\documentclass[twocolumn,showpacs,preprintnumbers,amsmath,amssymb,superscriptaddress]{revtex4}

\usepackage{graphicx}
\usepackage{bm}
\usepackage{color}

\usepackage{miller}
\usepackage{multirow}

\begin{document}

	\newcommand{\doped}{$\mathrm{(Nd_{1-x}Ce_x)_2Fe_{14}B}$}
\newcommand{\dopedco}{$\mathrm{(Nd_{1-x}Ce_x)_2Fe_{14-y}Co_yB}$}
\newcommand{\neo}{$\mathrm{Nd_2Fe_{14}B}$}
\newcommand{\dopedeleven}{$\mathrm{(Nd_{1-x}Ce_x)_2Fe_{14}^{11}B}$}
\newcommand{\isoneo}{$\mathrm{Nd_2Fe_{14}\!^{11}B}$}
\newcommand{\isoneotitle}{$\bm{\mathrm{Nd_2Fe_{14}\!^{11}B}}$}
\newcommand{\titleneo}{$\bm{\mathrm{Nd_2Fe_{14}B}}$}
\newcommand{\neuneo}{$\mathrm{(Nd_{0.78}Ce_{0.22})_2Fe_{14}B}$}
\newcommand{\isoneuneo}{$\mathrm{(Nd_{0.78}Ce_{0.22})_2Fe_{14}\!^{11}B}$}
\newcommand{\isosmallneo}{$\mathrm{(Nd_{0.91}Ce_{0.089})_2Fe_{14}\!^{11}B}$}
\newcommand{\isobigneo}{$\mathrm{(Nd_{0.62}Ce_{0.38})_2Fe_{14}\!^{11}B}$}
\newcommand{\bigneo}{$\mathrm{(Nd_{0.62}Ce_{0.38})_2Fe_{14}B}$}
\newcommand{\isoneuneoerr}{$\mathrm{\big( Nd_{0.78(4)}Ce_{0.22(1)}\big)_2Fe_{14}\!^{11}B}$}
\newcommand{\isosmallneoerr}{$\mathrm{\big(Nd_{0.91(8)}Ce_{0.089(7)}\big)_2Fe_{14}\!^{11}B}$}
\newcommand{\isobigneoerr}{$\mathrm{\big(Nd_{0.62(6)}Ce_{0.38(4)}\big)_2Fe_{14}\!^{11}B}$}
\newcommand{\bound}{$\mathrm{NdFe_4B_4}$}
\newcommand{\isobound}{$\mathrm{NdFe_4\!^{11}B_4}$}
\newcommand{\isoboundtiny}{$\mathrm{Nd_{0.904}Ce_{0.096}Fe_4\!^{11}B_4}$}
\newcommand{\isoboundtinyerr}{$\mathrm{Nd_{0.904(9)}Ce_{0.096(1)}Fe_4\!^{11}B_4}$}
\newcommand{\isoboundsmall}{$\mathrm{Nd_{0.75}Ce_{0.25}Fe_4\!^{11}B_4}$}
\newcommand{\isoboundsmallerr}{$\mathrm{Nd_{0.75(9)}Ce_{0.25(3)}Fe_4\!^{11}B_4}$}
\newcommand{\isoboundbig}{$\mathrm{Nd_{0.49}Ce_{0.51}Fe_4\!^{11}B_4}$}
\newcommand{\isoboundbigerr}{$\mathrm{Nd_{0.49(2)}Ce_{0.51(2)}Fe_4\!^{11}B_4}$}
\newcommand{\isoboundmed}{$\mathrm{Nd_{0.61}Ce_{0.39}Fe_4\!^{11}B_4}$}
\newcommand{\isoboundmederr}{$\mathrm{Nd_{0.61(3)}Ce_{0.39(2)}Fe_4\!^{11}B_4}$}
\newcommand{\dg}{$^{\circ}$}
\newcommand{\dc}{$\mathrm{^{\circ} C}$}
\newcommand{\mub}{$\mu_{B}$}
\newcommand{\ha}{$H_A$}
\newcommand{\ms}{$M_s$}
\newcommand{\tc}{$T_C$}
\newcommand{\ts}{$T_s$}
\newcommand{\hc}{$H_c$}
\newcommand{\br}{$B_r$}

\title{Growth and Characterization of Ce- Substituted \isoneotitle\ Single Crystals}

\author{M. A. Susner}

\email{susnerma@ornl.gov}

\author{B. S. Conner}
\affiliation{Materials Science and Technology Division, Oak Ridge National Laboratory, Oak Ridge, Tennessee 37831 USA}

\author{B. I. Saparov}
\affiliation{Materials Science and Technology Division, Oak Ridge National Laboratory, Oak Ridge, Tennessee 37831 USA}

\author{M. A. McGuire}
\affiliation{Materials Science and Technology Division, Oak Ridge National Laboratory, Oak Ridge, Tennessee 37831 USA}

\author{E. J. Crumlin}
\affiliation{Advanced Light Source, Lawrence Berkeley National Laboratory, Berkeley, California 94720 USA}

\author{G. M. Veith}
\affiliation{Advanced Light Source, Lawrence Berkeley National Laboratory, Berkeley, California 94720 USA}

\author{H. B. Cao}
\affiliation{Quantum Condensed Matter Division, Oak Ridge National Laboratory, Oak Ridge, Tennessee 37831 USA}

\author{K. V. Shanavas}
\author{D. S. Parker}
\affiliation{Materials Science and Technology Division, Oak Ridge National Laboratory, Oak Ridge, Tennessee 37831 USA}

\author{B. C. Chakoumakos}
\affiliation{Quantum Condensed Matter Division, Oak Ridge National Laboratory, Oak Ridge, Tennessee 37831 USA}

\author{B. C. Sales}

\affiliation{Materials Science and Technology Division, Oak Ridge National Laboratory, Oak Ridge, Tennessee 37831 USA}

\date{\today}

\begin{abstract}
Single crystals of \doped\ are grown out of Fe-(Nd,Ce) flux. Chemical and structural analysis of the crystals indicates that \doped\ forms a solid solution until at least $x = 0.38$ with a Vegard-like variation of the lattice constants with $x$.  Refinements of single crystal neutron diffraction data indicate that Ce has a slight site preference (7:3) for the $4g$ rare earth site over the $4f$ site. Magnetization measurements show that for $ x = 0.38$ the saturation magnetization at 400 K, a temperature important to applications, falls from 29.8 for the parent \neo\ to 27.6 \mub/f.u., the anisotropy field decreases from 5.5 T to 4.7 T, and the Curie temperature decreases from 586 to 543 K. First principles calculations carried out within density functional theory are used to explain the decrease in magnetic properties due to Ce substitution. Though the presence of the lower-cost and more abundant Ce slightly affects these important magnetic characteristics, this decrease is not large enough to affect a multitude of applications. Ce-substituted \neo\ is therefore a potential high-performance permanent magnet material with substantially reduced Nd content.
\end{abstract}


\maketitle

\section{Introduction}

The reduction of critical materials present in consumer and industrial products in general, and in Rare Earth (RE) -based permanent magnets in particular, is a pressing concern due to increasing demand and a decline and/or uncertainty in the supply chain of these materials. Dysprosium, judged to be the most critical of these elements \cite{_critical_2011}, is substituted onto the Nd site in \neo\ (2-14-1) permanent magnets to increase the value of their coercive field and maximum operating temperatures \cite{sagawa_permanent_1984, croat_praseodymium-iron-_1984}. Depending on the application, the amount of Dy added can be large, in the range of 1.4-8.7 wt\% \cite{_critical_2011}. The highest temperature applications, those that rely on the greatest quantities of dysprosium, are required by the rapidly emerging technologies of wind turbines, magnetically levitated transport, and traction motors for hybrid and electric cars \cite{_critical_2011}. 

Though Dy is indeed considered the most critical element, by many criteria Nd is not that far behind on the list \cite{_critical_2011}. It is therefore imperative that new and innovative solutions are applied to the synthesis of permanent magnets that contain more abundant elements. To this end, recent work by Pathak $et$ $al$. \cite{pathak_cerium:_2015} has shown that partial substitution of Ce for Nd and Co for Fe ($i.e.$ $ \mathrm{Nd_{1.6}Ce_{0.4}Fe_{12}Co_{2}B}$) results in a permanent magnet with properties superior to those of Dy-substituted 2-14-1 magnets for temperatures above 450 K. Since 1) Ce is by far the most abundant of the rare earth elements and 2) these new magnets simultaneously eliminate the need for Dy and reduce the amount of Nd used, they represent an exciting opportunity for the development of lower-cost high performance permanent magnets. 

The crystal structure of \neo\ was established as tetragonal (space group $P4_2/mnm$, No. 136) as early as 1984 by the authors of Refs. \cite{givord_magnetic_1984} and \cite{herbst_relationships_1984}. There are two separate and inequivalent RE sites; one is a $4g$ site and the other a $4f$ site. Also present are six inequivalent Fe sites and one unique B site. Past investigations as to the effects of chemical substitution of different RE elements onto the Nd site are numerous, see for example Ref. \cite{herbst_rare_1991} and the references therein. However, detailed, quantitative experimental studies on the effects of Ce substitution of varying amounts in pure single crystal samples have not previously been performed. Abache and Osterreicher \cite{abache_magnetic_1986} studied aligned powders to determine the crystal structure, magnetic anisotropy, and spin reorientation temperature in \doped\ for $x=0.25$ and $x=1$ along with a multitude of other compositional variations of \neo\ achieved by varying the degree and type of substitutions on the RE and transition metal (TM) sites. However, no detailed study was made concerning the magnetic properties as a function of Ce concentration in this work. They also do not make any claims whatsoever as to the stability of the \doped\ compound for $0.25 \leq x \leq 1$. More recently, density functional theory calculations of Alam $et$ $al$. \cite{alam_site-preference_2013} actually predict a phase segregation for $x \geq 0.3$ in \doped.  Pathak $et$ $al$. \cite{pathak_cerium:_2015}  have experimentally observed phase segregation under the conditions required for the fabrication of melt-spun ribbons.

Information also of interest in the \doped\ system is the site preference of the Ce as there are two structurally distinct RE sites. Previous theoretical predictions about Ce site preference exist \cite{abache_magnetic_1986, alam_site-preference_2013}, but direct experimental evidence to determine whether Ce in \doped\ is preferentially associated with the larger $4g$ site or the smaller $4f$ site has previously been lacking.

\begin{table}
  \centering
  \caption{Molar Ratios for Sample Fluxes and Compositions of Single Crystals as determined from EDS.}
\resizebox{\columnwidth}{!}{
    \begin{tabular}{cccccc}
  \\
    \multicolumn{4}{c}{\textbf{Flux Composition}} & &  \\
 
    \multicolumn{4}{c}{\textbf{(molar percent)}} &    \multicolumn{1}{c}{\textbf{Composition of }}  & \multicolumn{1}{c}{\textbf{Composition of }}   \\
       Nd    & Ce    & Fe    & $^{11}\mathrm{B}$    &      \multicolumn{1}{c}{\textbf{2-14-1 phase}}   &  \multicolumn{1}{c}{\textbf{1-4-4 phase}} \\
    \cline{1-6} \\
    51    &  0    & 43.3  & 5.8   & \isoneo & \isobound \\
    43.3  & 7.7   & 43.3  & 5.8   & \isosmallneoerr & \isoboundtinyerr\\
    32.9  & 18.1  & 43.3  & 5.8   & \isoneuneoerr & \isoboundsmallerr \\
    25.5  & 25.5  & 43.3  & 5.8   &\isobigneoerr & \isoboundmederr \\
    18.1  & 32.9  & 43.3  & 5.8   &  \hfill  \footnote{ 2-14-1 phase did not form large crystals}  \hfill & \isoboundbigerr \\

    \end{tabular}%
}
  \label{table1}%
\end{table}%

The \doped\ system has recently come under much attention. Experimental investigations by Pathak $et$ $al$. discovered an anomaly evident in the plot of the $c$ lattice parameter vs. Ce concentration in melt-spun \doped\ ribbons for $x \approx 0.2$ \cite{pathak_cerium:_2015}. Similar anomalies were evident in plots of coercivity (\hc), maximum energy product ($BH_{\mathrm{max}}$), and remnant magnetization (\br) as functions of Ce concentration, along with the aforementioned phase segregation. The same investigation yielded an extraordinary result: when Co is substituted for Fe by two atoms per formula unit in \dopedco\ where $ x = 0.20$, the Curie temperature (\tc) increased by 150 K while coercivity dropped only 22\% as compared to an equivalent sample with no Co doping, $ i.e.$ $x = 0.2, y = 0$. The anisotropy field (\ha), saturation moment (\ms), and \br\ remained more or less unchanged compared to the $ x = 0.2, y = 2 $ case. The possible applications of such excellent Dy-free magnets spawns an exciting line of research focused on linking the microscopic magnetic properties in pure crystalline samples with the macroscopic magnetic properties of bulk permanent magnets.

In the current work, we have employed single crystal growth techniques to synthesize single crystals of Ce-doped 2-14-1 compounds. We show here observations contrary to those of the melt-spun ribbons of Ref. \cite{pathak_cerium:_2015} for our samples grown under the slow crystallizing conditions of flux growth; we are able to stabilize uniform single crystals of \doped\ for values $ 0 \leq x \leq 0.38$. Furthermore, we have used these single crystal specimens to experimentally determine that Ce has a slight site preference in the 2-14-1 structure for the larger RE $4g$ site  \cite{abache_magnetic_1986, alam_site-preference_2013}. Next, we have also measured important magnetic properties such as \ms, \tc, \ha\ etc. for \doped\ for the composition range $ 0 \leq x \leq 0.38$. Finally, we conclude this report with first principles density functional theory calculations which help explain the causes of the changes in \ms, \tc, and \ha\ with Ce substitution. 

\section{Methods}

Single crystals of both \neo\ and \bound\ type materials were grown from Nd-Fe flux following the techniques originally reported by Canfield $et$ $al$. \cite{canfield_high-temperature_2001, canfield_private_2014, lewis_magnetic_1998} and further refined by  Saparov \cite{saparov_private_2014}. The starting materials were cuttings of high purity metals: Ce (Ames Laboratory, 99.99\%), Nd (Ames Laboratory, 99.99\%), Fe (Alfa Aesar, 99.98\%), Co (Alfa Aesar, 99.95\%) and isotopically pure $\mathrm{^{11}B}$ ($\mathrm{^{11}B}$, ORNL, 99.99+\%). The use of $\mathrm{^{11}B}$ was necessary for neutron diffraction experiments as $\mathrm{^{10}B}$ is exceptionally good at neutron capture. Appropriate stoichiometries of these elements (see Table \ref{table1}) were loaded into Ta crucibles (1.25 cm diameter, 7 cm length) and sealed under ~0.5 atm Ar using an arc-melter. A Ta frit was placed above the starting materials to act as a filter during the centrifugation process. The Ta crucibles were subsequently sealed in quartz ampoules under $\sim$1/3 atm Ar. 

\begin{figure}[h]
\centering
\includegraphics[width=0.48\textwidth, trim={2cm 5.5cm 1cm 6.5cm },clip]{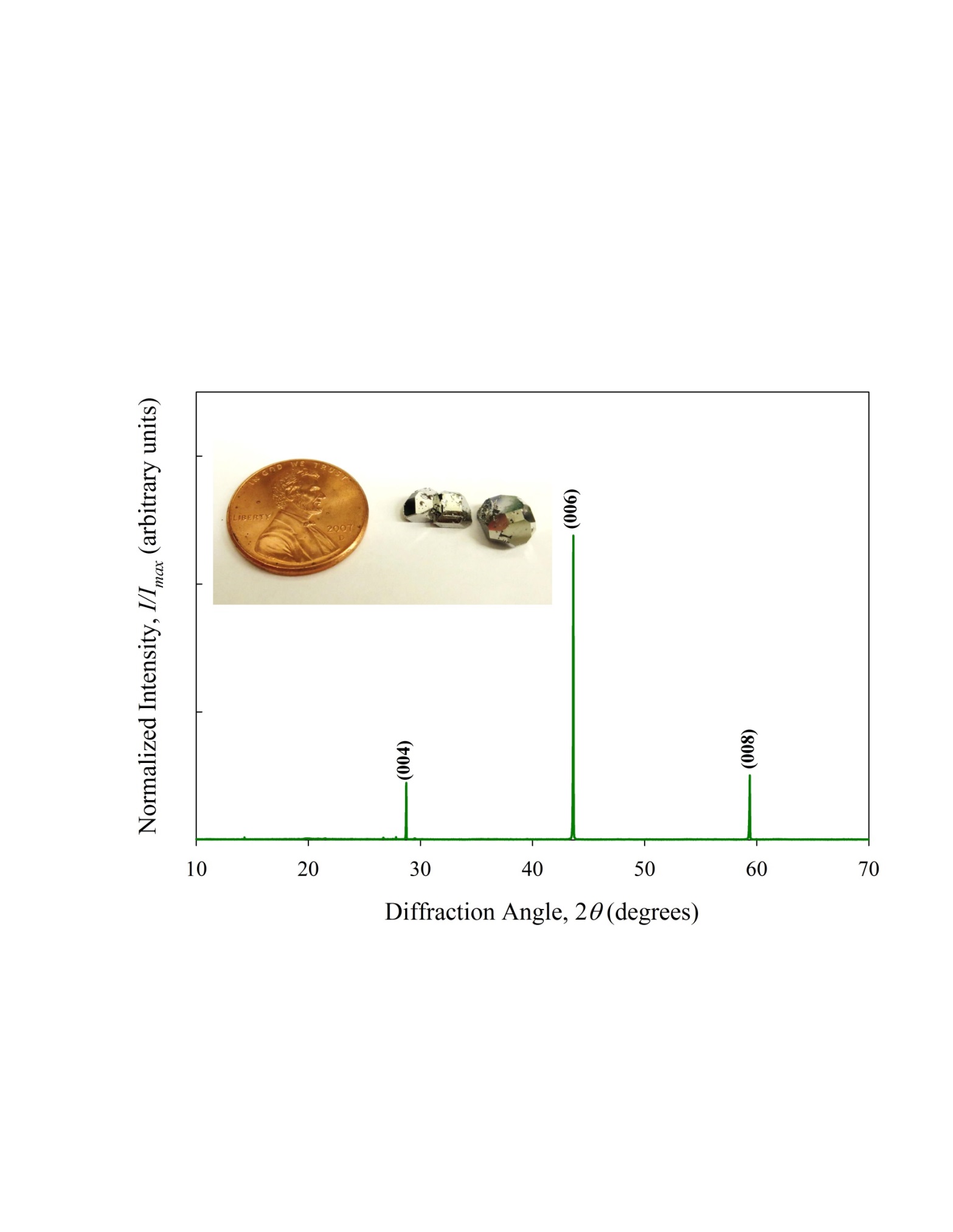}
\caption{XRD diffraction pattern from \hkl(0 0 1) face of undoped \neo\ crystal. The inset shows a size comparison image of crystals of the same composition. }
\label{figure1}
\end{figure}

The sealed ampoules were placed into a large box furnace and heated to 1190  \dc\ over 12 h and held at that temperature for 24 h. The furnace was then cooled to 800 \dc\ over 390 h, after which the samples were removed and the flux was decanted using a centrifuge. Large, single crystal specimens (some on the order of $\sim$1 g) of the \neo\ (2-14-1) phase (Fig. \ref{figure1}, inset) and smaller elongated prisms ($\sim$ 2 mm $\times$ 2 mm $\times$ 10 mm) of the \bound\ (1-4-4) phase were extracted from the crucibles. The ultimate sizes of the 2-14-1 crystals were not influenced by the compositions of the fluxes with the exception of the largest Ce concentration (in this case, large crystals of the Laves phase compound $\mathrm{Ce_{0.85}Nd_{0.15}Fe_2}$ formed rather than the 2-14-1 phase). The crystallographic faces of all phases present after decanting were well-defined. In Fig. \ref{figure1} we present XRD reflections resulting from an \hkl(0 0 1) face of the undoped \neo\ sample.

Analyses of elemental compositions were performed using a Hitachi TM-3000 electron microscope equipped with a Bruker Quantax 70 energy dispersive (EDS) X-ray system ($cf.$ Table \ref{table1}). For all compositions small crystals were ground into powder and used to produce XRD patterns using a PANalytical X’Pert Pro diffractometer (Cu $K_{\alpha}$, 1.54056 $\AA$). The resultant peak reflections were analyzed for lattice parameters using the LeBail fitting function available with the onboard HighScore Plus software package. Single crystal diffraction was performed using a Rigaku single crystal X-ray diffractometer with Dectris Pilatus 200K detector (Mo $K_{\alpha}$, 0.71073 $\AA$); the resultant diffraction patterns were analyzed using the onboard Rigaku software with absorption corrections performed using a spherical approximation. The structure was refined using SHELXL (Ref. \cite{sheldrick:sc5010}) with WinGX. Single crystal neutron diffraction was measured using the HB-3A four circle diffractometer at the High Flux Isotope Reactor (HFIR) at the Oak Ridge National Laboratory. The $\sim$300 mg sample was measured at 4 K and 300 K by neutrons with a  wavelength of 1.003 $\AA$ from a bent Si-331 monochromator \cite{Chako_2011}. The structure was refined using the FULLPROF software package \cite{carvajal_1993}.

X-ray photoelectron spectroscopy (XPS) data were collected with a PHI 3056 XPS spectrometer with an Al $K_{\alpha}$ source (1.4866 keV) in a cryo-pumped ultra-high vacuum chamber with a pressure of  $<< 10^{-8}$ Torr. Fresh surfaces of the crystal were created by grinding a single crystal in an Ar-filled glove box and transferring to the XPS chamber using a vacuum transfer system. Synchrotron data was taken at the Advanced Light Source (ALS) at the Lawrence Berkeley National Laboratory (Berkeley, USA) using Beamline 9.3.1. This beam-line is a bent magnet beam-line with an energy range of 2.3$-$5.2 keV. The minimal spot size at the beam-line is 0.7 mm (v) $\times$ 1.0 mm (h) \cite{axnanda_using_2015}. 

Magnetic properties were measured using 1) a Quantum Design Magnetic Property Measurement System (MPMS) and 2) the DC-extraction capability of the AC Measurement System (ACMS) option of a 14 T Quantum Design Physical Property Measurement System (PPMS). To measure the magnetization along the easy direction as a function of magnetic field 1) a $c$ axis face was found on a single crystal using the X-ray diffractometer, 2) the crystal was polished into a parallelepiped with a long axis aligned with either the $a$ or $c$ direction (for hard and easy axis measurements, respectively) and 3) the crystal segments were placed into the MPMS or PPMS under magnetic fields of up to 13 T.  Due to the tendency for the easy axis to align with the applied field direction, single crystal specimens were unable to quantitatively yield correct hard axis magnetization measurements. To obtain these measurements, small crystals of each composition were ground into powder and placed into a gel cap with epoxy; the epoxy was allowed to set under an applied magnetic field so as to align the powders. Measurements were taken with $H$ oriented perpendicular to the easy direction at 5 K, 300K, 350 K, and 400 K. To obtain a correct value for the saturation magnetization a measurement of the same gel cap was taken with $H$ oriented parallel to the easy direction; these data were then normalized to the values obtained from the single crystal easy axis measurements. The easy axis measurements of the aligned powder samples and the single crystal specimens were in good agreement. 

The $M(T)$ properties of the 2-14-1 crystals were also measured using the Quantum Design MPMS. The magnetization was measured as a function of temperature using an applied magnetic field of $10^3$ Oe; temperature ranges were 2 K $ \leq T \leq$ 750 K. Further confirmation of the Curie temperatures of the grown ferromagnetic crystals was obtained through thermal analysis using a Perkin Elmer Pyris Diamond Thermo-Gravimetric Analyzer/ Differential Thermal Analyzer (TGA/DTA) with an applied magnetic field ($<$ 0.1 T).

 The first principles calculations are carried out within density functional theory (DFT). We use the generalized gradient approximation \cite{Perdew_1996} within the projector augmented wave method \cite{bloechl_projector_1994} as implemented in the Vienna ab initio simulation package \cite{Kresse_ab_1993, Kresse_efficient_1996}.  An energy cutoff of 400 eV and $k$ space sampling on a $7\times7\times5$ grid are employed. The Ce substitution on the Nd site is handled by the virtual crystal approximation (VCA).

\begin{table}
\begin{centering}
  \caption{Summary of 250 K single crystal X-ray diffraction data of sample \neuneo; \newline Space group $P4_2/mnm$; $a=b=8.8032(13) \AA$; $c=12.1880(20) \AA$; $\alpha=\beta=\gamma=90^{\circ}$}
\resizebox{\columnwidth}{!}{
    \begin{tabular}{lccccc}
& & & & & \\
 \multicolumn{1}{c}{} & \multicolumn{3}{c}{\textbf{Atomic Coordinates}} & \multicolumn{1}{c}{} \\
\multicolumn{1}{c}{\textbf{Site}} & x     & y     & z     & \multicolumn{1}{c}{$\bm {U_{iso}} (\AA^2)$} \\
\cline{2-4} \\
 B1  ($4g$) & 0.1239(8) & 0.1239(8) & 0     & 0.0121(16) \\
  Fe1 ($16k_1$) & 0.03732(7) & 0.35986(7) & 0.32411(5) & 0.0099(2) \\
Fe2 ($16k_2$) & 0.06698(7) & 0.27578(7) & 0.12753(5) & 0.099(2) \\
           Fe3 ($8j_1$) & 0.09802(7) & 0.09802(7) & 0.29599(7) & 0.0105(2) \\
     Fe4 ($8j_2$) & 0.31758(7) & 0.31758(7) & 0.25407(7) & 0.0103(2) \\
           Fe5 ($4e$) & 0     & 0     & 0.11514(10) & 0.0098(3) \\
           Fe6 ($4c$) & 0     &0.5     & 0     & 0.0105(3) \\
           Nd1/Ce1 ($4g$) & 0.22992(4) & 0.77008(4) & 0     & 0.0108(2) \\
          Nd2/Ce2 ($4f$) & 0.35743(4) & 0.35743(4) & 0     & 0.0108(2) \\
                 &       &       &       &  \\
 \cline{1-5} \\
  \multicolumn{1}{c}{\textbf{Reliability}} & \multicolumn{3}{c}{} & \multicolumn{1}{c}{} \\
  \multicolumn{1}{c}{\textbf{factors}} & \multicolumn{1}{c}{$R_1$}  & \multicolumn{1}{c}{$wR_2$}  & \multicolumn{1}{c}{$R_{int}$} & \multicolumn{1}{c}{GOF} \\
          & 0.0505 & 0.1189 & 0.0936 & 0.882   \\
   
    \end{tabular}%
}
  \label{table2}%
  \end{centering}
\end{table}%

\begin{table}
  \centering
  \caption{Summary of 250 K single crystal X-ray diffraction data of sample \neuneo; \newline Space group $P4_2/mnm$; $a=b=8.8041(13) \AA$; $c=12.1633(17) \AA$; $\alpha=\beta=\gamma=90^{\circ}$}
\resizebox{\columnwidth}{!}{
    \begin{tabular}{lccccc}
 \\
 \multicolumn{1}{c}{} & \multicolumn{3}{c}{\textbf{Atomic Coordinates}} & \multicolumn{1}{c}{} \\
\multicolumn{1}{c}{\textbf{Site}} & x     & y     & z     & \multicolumn{1}{c}{$\bm {U_{iso}} (\AA^2)$} \\
\cline{2-4} \\
 B1 ($4_g$) & 0.1221(14) & 0.1221(14) & 0     & 0.008(3) \\
   Fe1 ($16k_1$) & 0.03717(14) & 0.36003(14) & 0.32427(10) & 0.0093(4) \\
 Fe2 ($16k_2$) & 0.06714(14) & 0.27523(14) & 0.12785(11) & 0.0102(4) \\
 Fe3 ($8j_1$) & 0.09839(14) & 0.09839(14) & 0.29627(16) & 0.0108(4) \\
 Fe4 ($8j_2$) & 0.31744(13) & 0.31744(13) & 0.25413(14) & 0.0102(4) \\
 Fe5 ($4_e$) & 0     & 0     & 0.1155(2) & 0.0104(5) \\
 Fe6 ($4_c$) & 0     &0.5     & 0     & 0.0101(6) \\
 Nd1/Ce1 ($4g$) & 0.23015(7) & 0.76985(7) & 0     & 0.0102(3) \\
 Nd2/Ce2 ($4f$) & 0.35750(7) & 0.35750(7) & 0     & 0.0102(3) \\
              \\
    \cline{1-5} \\
  \multicolumn{1}{c}{\textbf{Reliability}} & \multicolumn{3}{c}{} & \multicolumn{1}{c}{} \\
  \multicolumn{1}{c}{\textbf{factors}} & \multicolumn{1}{c}{$R_1$}  & \multicolumn{1}{c}{$wR_2$}  & \multicolumn{1}{c}{$R_{int}$} & \multicolumn{1}{c}{GOF} \\
          &  0.0792 & 0.1849 & 0.0964 & 1.196 &   \\
    \end{tabular}%
}
  \label{table3}%
\end{table}%

 \section{Results and Discussion}
 \subsection{Crystal growth and Ce doping}

	The as-grown crystals were roughly cubic in shape and displayed well-defined faceting (Fig. \ref{figure1}, inset). The sizes of the 2-14-1 crystals were typically $\sim2\times2\times2$ mm$^3$ with some reaching $\sim10\times10\times10$ mm$^3$ in volume. Except for the growth with a nominal flux composition of 18.1\% Nd and 32.9\% Ce where a Laves phase compound formed rather than the 2-14-1 phase, all resulting single crystals were of similar size. The 1-4-4 phases grew as needles $\sim 1\times1\times10$ mm$^3$ and will be the subject of a forthcoming publication.

Through EDS analysis, we were able to measure the relative concentrations of the rare earths (RE) and transition metals (TM). The boron content could not be reliably estimated from the EDS measurements. A variety of growths were used to maximize the Ce concentrations within the crystals. Results suggest that $ x = 0.38$ is the maximum attainable by this synthesis route.

\subsection{Structural characterization}

\begin{figure}[h]
\centering
\includegraphics[width=0.48\textwidth, trim={1cm 6cm 5cm 7cm},clip]{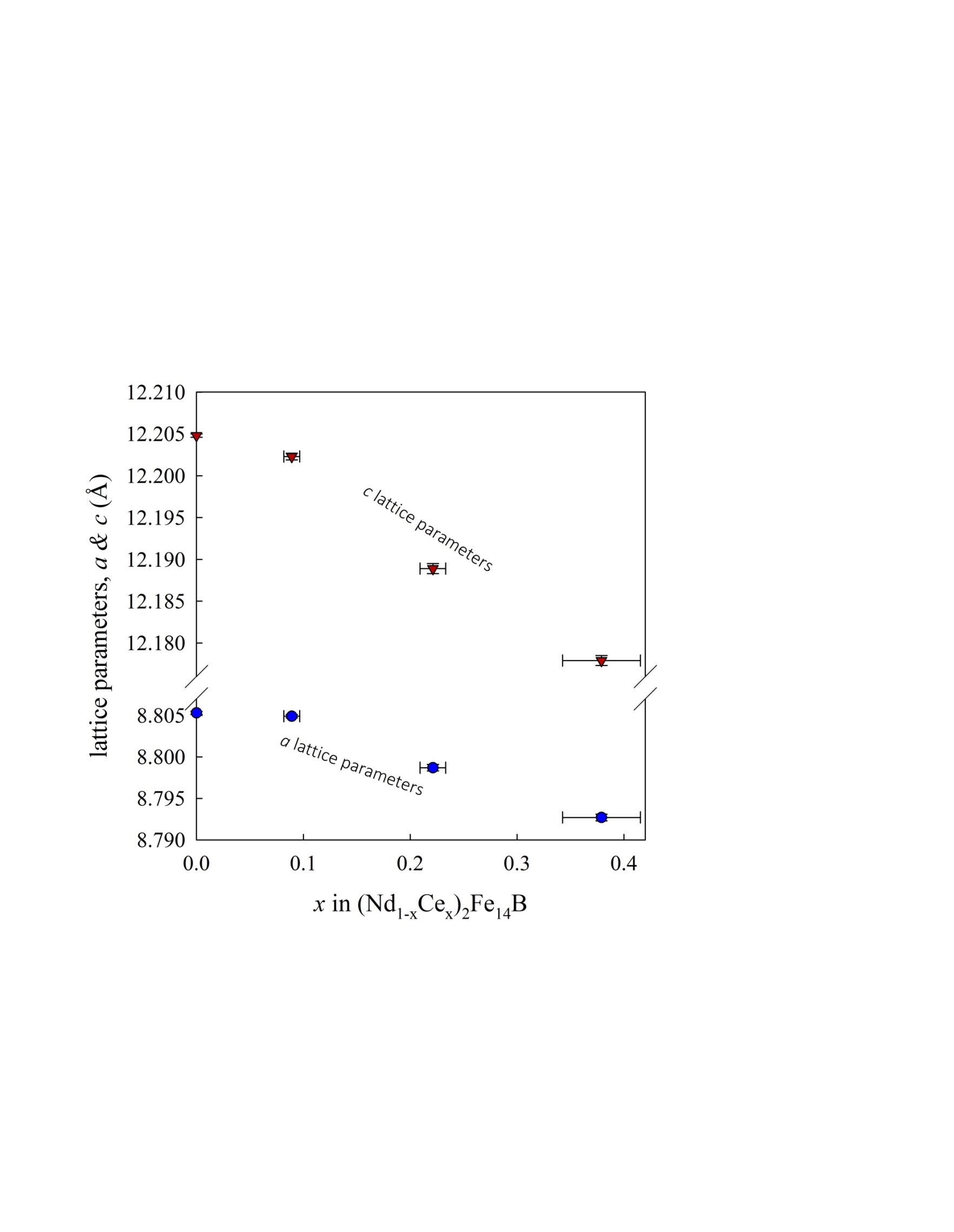}
\caption{Lattice parameters of the 2-14-1 phase plotted against Ce concentration. }
\label{figure2}
\end{figure}

The room temperature powder XRD patterns were taken of the 2-14-1 phases using the powder from ground crystals. LeBail fitting was used to extract the lattice parameters, which are plotted in Fig. \ref{figure2} as functions of Ce concentration. The values of both the $a$ and $c$ lattice parameters decrease monotonically with increasing quantities of Ce ($cf.$ Fig. \ref{figure2}).  At $x = 0.38$ in \doped\ we see that the value of a is 0.14\% smaller than that of the undoped sample. Similarly, for the same composition, $c$ is 0.22\% smaller. No anomalies are present in the lattice parameters as a function of Ce content, in contrast with the sintered magnets of the same composition synthesized by Pathak $et$ $al$. \cite{pathak_cerium:_2015} where a two-phase region was observed for $0.15 < x < 0.4$ in \doped. However, it should be noted that there is a wide spacing between our data points which could potentially belie small anomalies in the lattice parameters as a function of Ce content.  Our single crystal specimens containing the two highest compositions of Ce ($x = 0.22$ and $x = 0.38$) fall within the two phase region suggested by the above authors, indicating that the range of solid solution in this system is sensitive to processing conditions such as temperature and solidification rate.

\begin{figure}[h]
\centering
\includegraphics[width=0.45\textwidth]{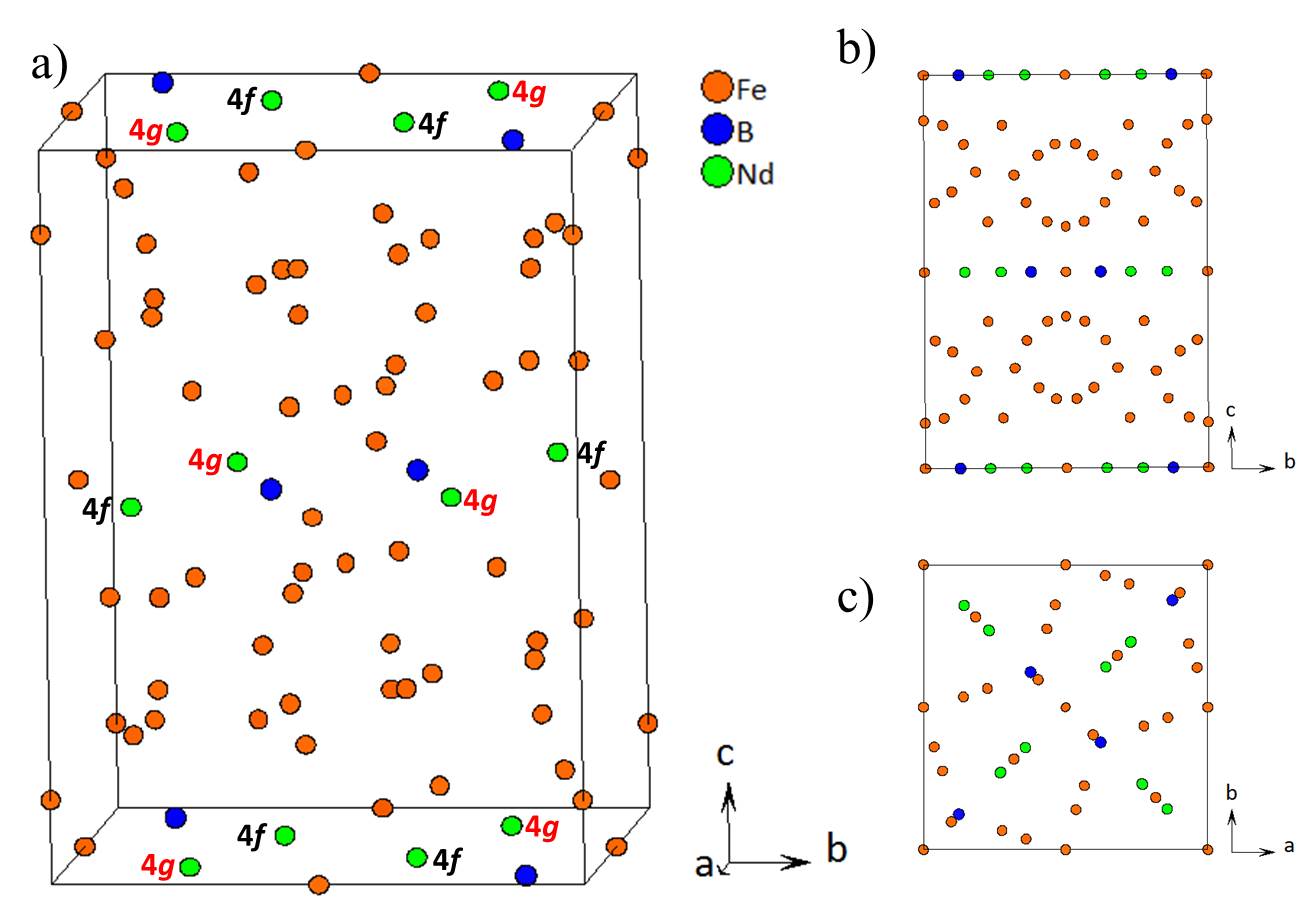}
\caption{250 K structural refinement of sample \neuneo: a) isometric view showing the positions of the two different RE sites: $4g$ and $4f$ (the larger $4g$ site is denoted by a red label); b) view along a axis; c) view along c axis.}
\label{figure3}
\end{figure}

Single crystal X-ray diffraction on the sample with composition \neuneo\ was performed at 250 K and 110 K; these temperatures bracket the spin reorientation temperature of $\sim$140 K in \neo\ \cite{abache_magnetic_1986, abache_magnetic_1985}. Structural refinements were made using SHELX \cite{sheldrick:sc5010} from 1068 and 966 unique reflections, respectively, and are presented in Tables \ref{table2} and \ref{table3}.  The space group matched that of previous reports and the powder diffraction data, $P4_2/mnm$. The values from the single crystal data match fairly well with those obtained from powder diffraction. No discernible structural difference was seen between the data collected above the spin reorientation temperature and that collected below it. For the single crystal analyses no attempts were made to refine the RE position in terms of Nd/Ce occupancy as the scattering factors of the two RE elements were too similar. Later refinements using the occupancy results elucidated from the neutron diffraction data (below) yielded no change in the refinement. The structure derived from the 250 K single crystal refinement is presented in Figs. \ref{figure3}a$-$\ref{figure3}c and matches well with those published in the literature \cite{herbst_relationships_1984, herbst_rare_1991, herbst_preferential_1986, dong_2015, teplykh_2013}.

Single crystal neutron diffraction was performed on a $\sim$300 mg sample of \neuneo\ at temperatures of 300 K and 4 K. We collected $\sim$630 reflections at each temperature. The magnetic signals were found at both 4 K and 300 K with a propagation vector of $k = 0$, i.e., they coincide with the nuclear peaks. Therefore, in the data refinements, the nuclear and magnetic structure parameters were refined together. The results are presented in Tables \ref{table4} and \ref{table5}. 

Given the fact that the easy axis of this material is along the \hkl[0 0 1] direction, the magnetic vectors at 300 K were constrained as such to reduce the total number of refined parameters. At 300 K the total refined magnetic moment is $36.76 \pm 2.32$  \mub/f.u., in reasonable accord with the MPMS-determined value of 32.3 \mub/f.u. The Fe sites have moments ranging from 1.3(1)$-$3.2(2) \mub. This variability is likely caused by the different steric environments experienced by each of the different Fe sites. With the singular exception of the $4e$ Fe site the size of the moments are correlated to the degree of Fe coordination around the Fe atoms. For example, two of the largest moments are seen on the $8j_2$ and $16k_1$ sites which have Fe coordinations of 12 and 10, respectively. The smallest moment is observed at the Fe $4c$ site where the highest RE coordination is present, with 4 of the 12 nearest neighbors being Nd/Ce. The two RE sites have moments of 2.372 and 1.605 \mub\ for $4g$ and $4f$, respectively.  The $4g$ site is, in terms of size, the larger of the two and is therefore predicted to be preferred for Ce, assuming the Ce is in the +3 oxidation state \cite{alam_site-preference_2013}. The Ce$^{3+}$ ion (115 pm diameter) is larger than the Ce$^{4+}$ ion (101 pm) and is therefore likely to prefer this larger site. When requiring that the thermal displacement parameters for the two different rare earth sites be equal, our 300K refinements presented in Table \ref{table4} show that this preference is not absolute in that Ce site preference for the $4g$ site is 70\%. If, however, we let the atomic displacement parameters for each of the RE sites refine independently, the preference of Ce for the $4g$ site becomes even more absolute at 86\%. Though the degree of Ce preference for the $4g$ site and the atomic displacement parameters are highly correlated in these refinements, the fact that Ce does prefer the $4g$ site does not change.  These observations concur with the preliminary XPS data taken at the Advanced Light Source (ALS) at the Lawrence Berkeley National Laboratory whereby the XPS spectrum for the Ce $3d$ electrons most closely resembles that of $\mathrm{Ce_2O_3}$, indicating that the majority of Ce is in the +3 oxidation state for this composition. Further experimentation is necessary to confirm this initial experimental result.

For the 4 K refinement, we fixed the Ce occupancy to the values found at 300 K (70\% on the RE $4g$ site and 30\% on the RE $4f$ site) and let the magnetic vectors vary in direction, as below $\sim$140 K the easy axis is no longer the \hkl[0 0 1] direction. At 4 K, the total magnetic moment for the sample is refined to be $49.1 \pm 10.7$ \mub in reasonable agreement with the experimentally determined value of 34.9 \mub/f.u. for this composition (below). The Fe sites have moments that refine in the range $2.819-3.528 \ \mu_B$. Again, the moments associated with the Fe sites largely scale with their degrees of coordination by other Fe atoms. The refinement also shows that at 4 K the moments are tilted by $\sim$39.0\dg,  consistent with the well-known phenomena of spin reorientation in the 2-14-1 compounds at  $\sim$140 K \cite{abache_magnetic_1986, abache_magnetic_1985} whereby the magnetic moments orient themselves along the \hkl[3 3 5] direction rather than the \hkl[0 0 1] direction seen at room temperatures \cite{givord_polarized_1985}. The angle between these two directions is 49.6\dg, varying slightly from the experimentally determined value of 39.0\dg\ from our refinement. However, to date no work has been done to quantify the effect of Ce substitution on this reorientation with respect to changes in the easy direction of magnetization.

\subsection{Spin re-orientation temperature}

\begin{figure}[h]
\centering
\includegraphics[width=0.45\textwidth]{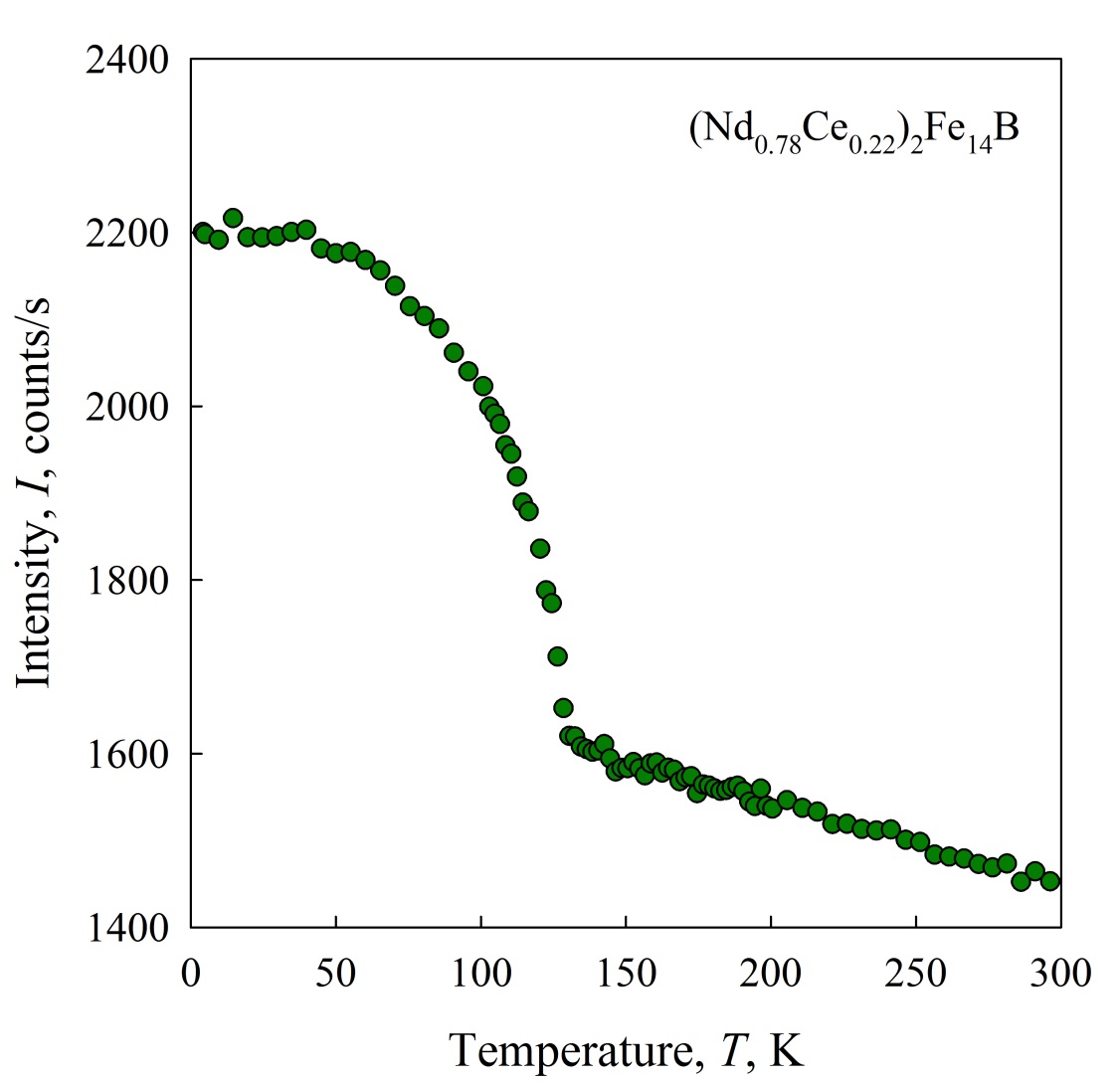}
\caption{Changes in (006) reflection intensity as measured from 4-300 K using single crystal neutron diffraction on sample with composition \isoneuneo. }
\label{figure4}
\end{figure}

Using single crystal neutron diffraction, the \hkl(0 0 6) reflection of \doped\ was measured upon warming (Fig.\ref{figure4}). Referring to Fig. \ref{figure4}, we see at 130.5 K a noticeable increase in the \hkl(0 0 6) peak intensity is evident, representing the spin reorientation transition whereby the magnetization easy axis shifts from the \hkl[0 0 1] direction at high temperatures to something near the \hkl[3 3 5] direction at low temperatures. Previous investigations of undoped \neo\  by Abache and Oesterreicher \cite{abache_magnetic_1986, abache_magnetic_1985} attributed this re-orientation to competing anisotropies of the $4g$ and $4f$ RE sites. In their argument, the $4f$ site (the smaller of the two) is more susceptible to planar moment alignment. At temperatures above the spin re-orientation temperature (\ts) the axial anisotropies of the RE and TM sites force the $4f$ site to align axially in a ferromagnetic structure. However, below 140 K the preference of the $4f$ site to align its moment in a planar orientation is much stronger compared to the competing influences of the Fe sites, resulting in spin re-orientation and a new easy axis for the system about 49\dg\ canted away from the $c$ axis \cite{abache_magnetic_1986, abache_magnetic_1985}. The spin re-orientation is therefore dependent on the $4f$ site. The related compound $\mathrm{Ce_2Fe_{14}B}$ exhibits no spin re-orientation \cite{abache_magnetic_1986} implying that Ce has no inclination for planar orientation of its moment. Substituting Ce on the $4f$ site is expected to affect the spin-reorientation. 

\begin{figure}[h]
  \includegraphics[width=.9\columnwidth, trim={0cm 0cm 0cm 1.5cm}]{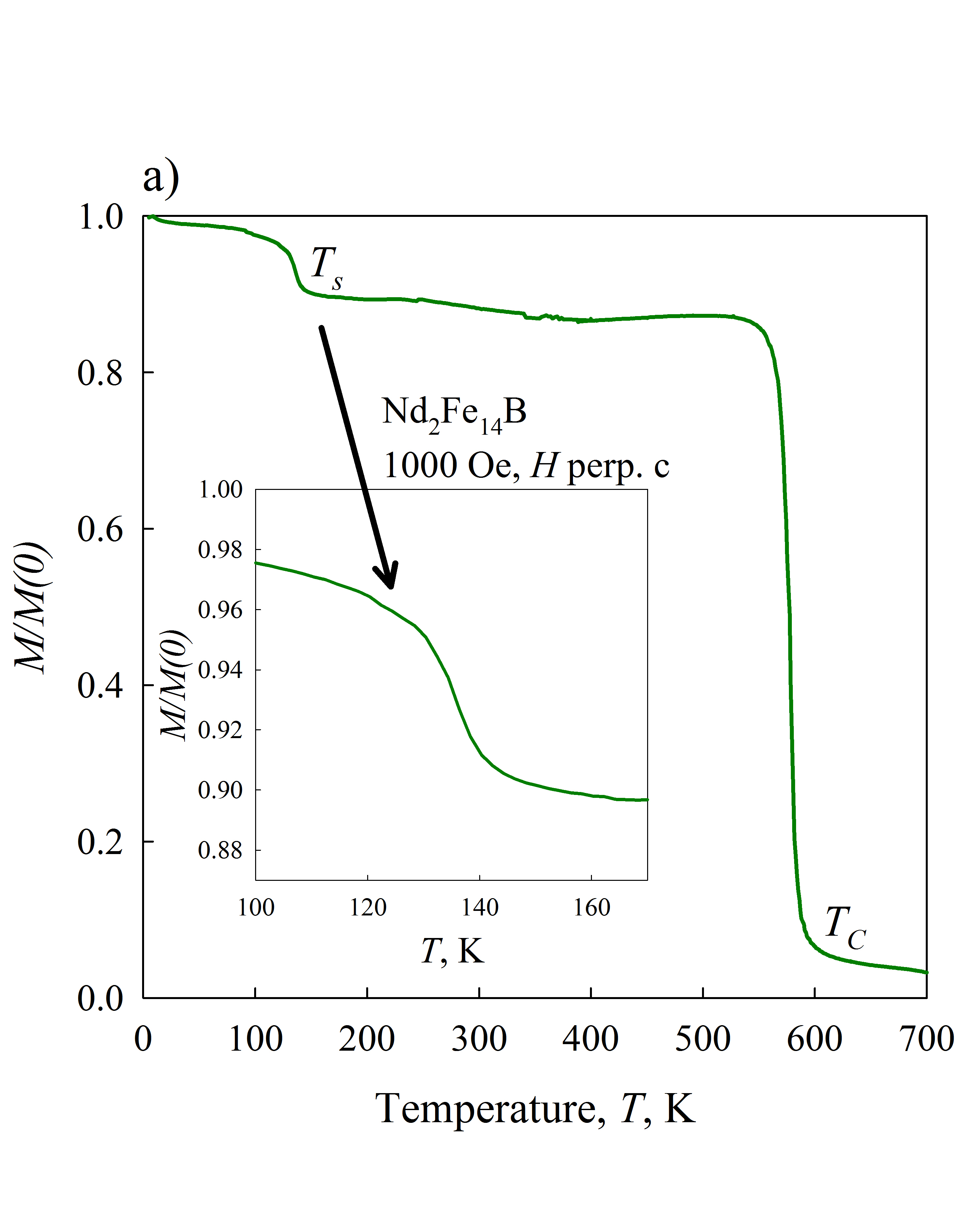}%

  \includegraphics[width=.9\columnwidth, trim={0cm 0cm 0cm 0.5cm}]{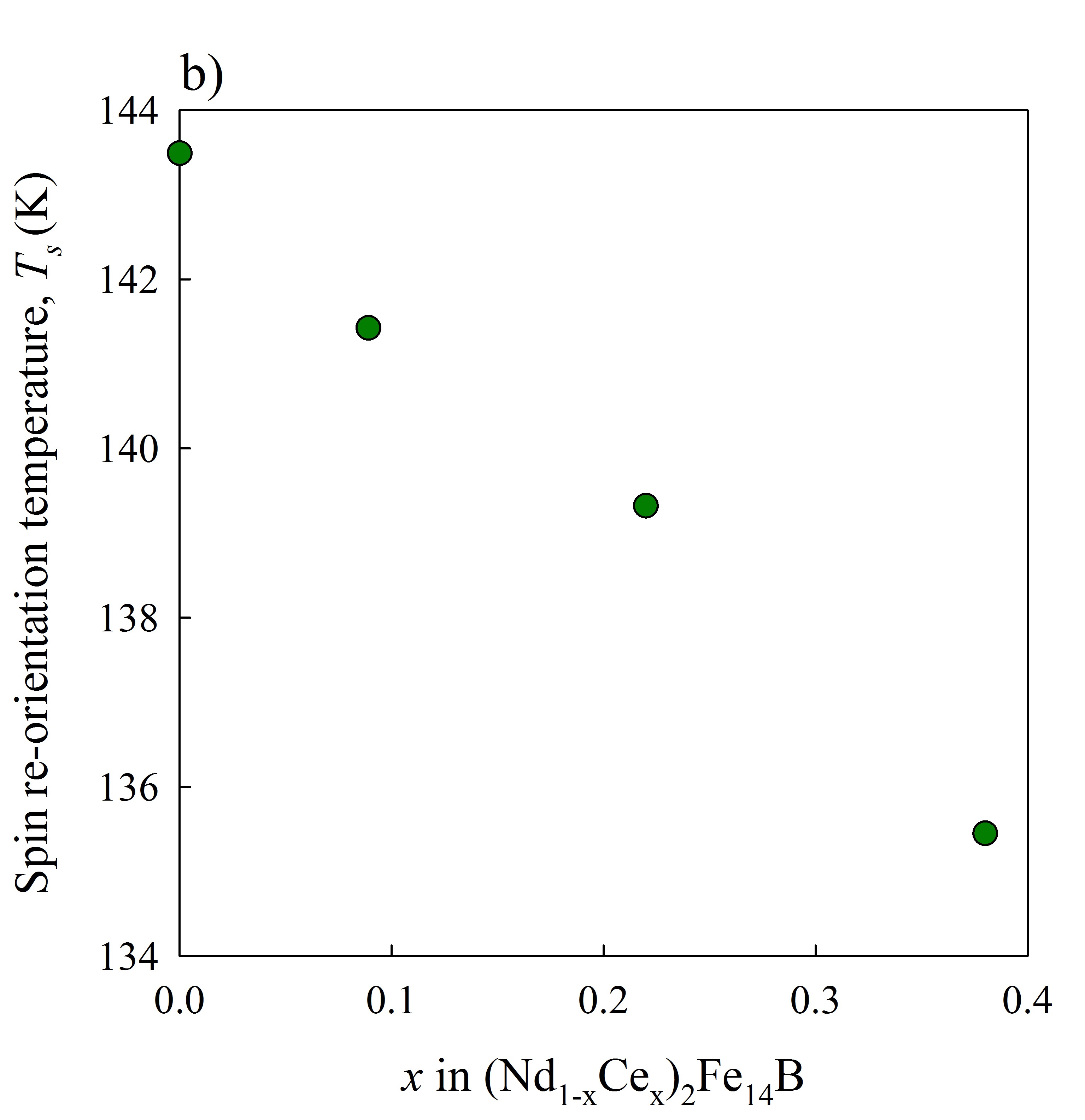}%
\caption{a) Normalized Magnetization as a function of temperature for the undoped \neo\ sample taken using 1000 Oe showing both the Curie and spin re-orientation temperatures (\tc\ and \ts, respectively) and b) Evolution of spin reorientation temperature as a function of Ce concentration.}
\label{figure5}
\end{figure}

In Fig. \ref{figure5}a we plot the magnetization of undoped \neo\ as a function of temperature and see increases in magnetization at 586 K and 143.5 K associated with \tc\ and \ts, respectively. In Fig. \ref{figure5}b we plot the evolution of \ts\ as a function of Ce doping. From the neutron diffraction refinements we know that ~18\% of the RE $4f$ site is substituted by Ce for the sample with the composition \neuneo. Presumably similar levels of Ce are found at this site for the other Ce compositions investigated concomitant with the level of Ce introduced. We see from Fig. \ref{figure5}b that \ts\ is sensitive to the amount of Ce present, indicating that although Ce does prefer the $4g$ site a not insignificant amount of Ce is still substituting onto the $4f$ site, assuming that the only way to modify the spin re-orientation temperature is to alter the occupancy of the $4f$ site. This contrasts with the prediction of Alam $et$ $al$. whereby Ce was predicted to sit on only the $4g$ site \cite{alam_site-preference_2013}.

\subsection{Magnetic properties: Curie temperature, anisotropy field, and saturation magnetization}

Relevant magnetic properties are summarized in Table \ref{table4}. The Curie temperatures of the Ce-substituted 2-14-1 samples were measured via MPMS and DTA/TGA. A permanent magnet affixed to the exterior of the sample chamber were used to produce magnetic field gradient at the sample, so that a change in magnetic susceptibility produced a change in the apparent sample mass. The data from these analyses show a steady monotonic decrease in \tc\ with increasing Ce concentration. The undoped \neo\ yields a \tc\ of $\approx$ 584 K. The composition \isobigneo, on the other hand, shows a \tc\ of $\approx$ 544 K, a total decrease of 1.1\% Ce substituted. This result is less than the rate one would expect if a Vegard-like relation existed between the \neo\ and $\mathrm{Ce_2Fe_{14}B}$ (\tc\ $\approx$ 430 K \cite{abache_magnetic_1986,abache_magnetic_1985}) parent compounds.

\begin{figure}[h]
  \includegraphics[clip,width=.9\columnwidth, trim={1.5cm 4cm 4cm 3.8cm}]{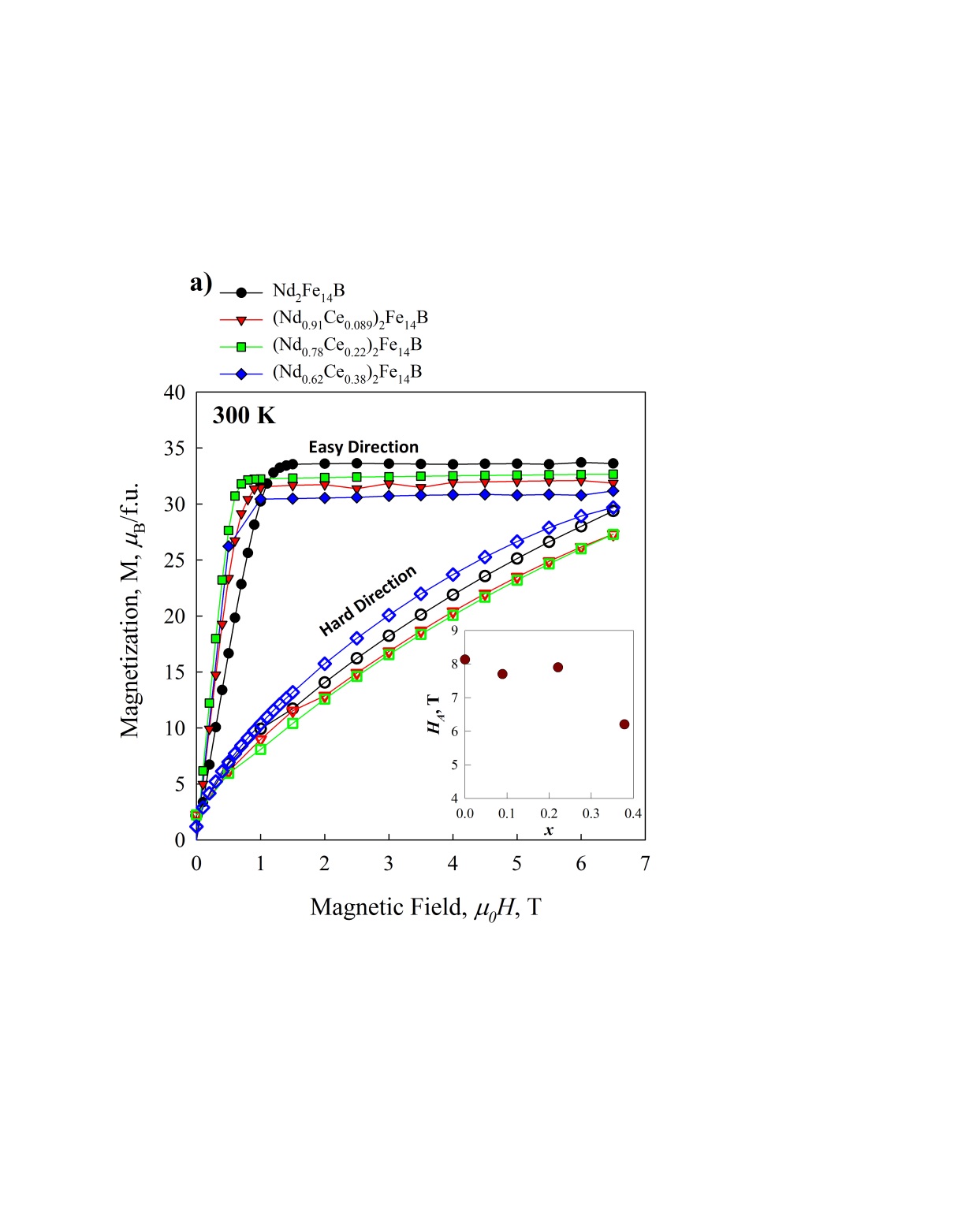}%
  
  \includegraphics[clip,width=.85\columnwidth, trim={1.5cm 4cm 4cm 3.8cm}]{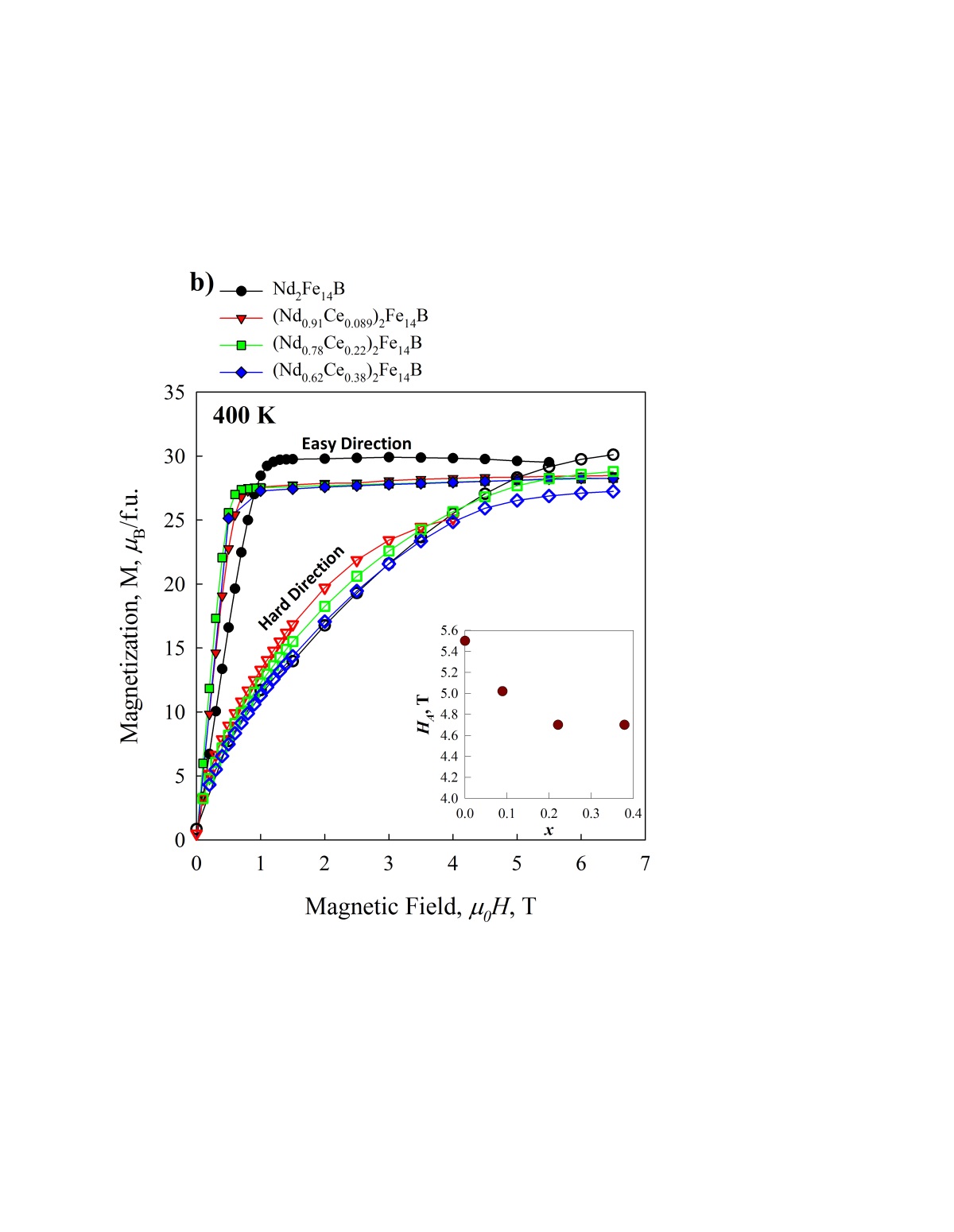}%
\caption{M(H) curves for undoped and Ce-substituted \neo\ samples at: a) 300 K and b) 400 K. The insets display the anisotropy fields, \ha, as functions of doping. }
\label{figure6}
\end{figure}

In Figs. \ref{figure6}a and \ref{figure6}b we show the $M(H)$ curves for the four Ce-substituted samples at 300 K and 400 K, respectively. With 38\% Ce doping, the saturation magnetization, Ms, decreases 9.2\% at 300 K (30.5 \mub/f.u. for \bigneo\ vs. 33.6 \mub/f.u. for the undoped \neo). Comparing these values to the 300 K saturation magnetizations of the parent compounds (23.9 \mub/f.u. for $\mathrm{Ce_2Fe_{14}B}$ \cite{hirosawa_magnetization_1986}) we find that the value of \ms\ seems to linearly scale with composition.  The effects at higher temperatures are similar. At 400 K, for the same composition \ms\ is decreased by 7.4\% from the undoped value. Concomitant with the decreases in \ms\ with increasing Ce concentration are decreases in \ha\, as determined from extrapolating the hard axis magnetization curves in Figs. \ref{figure6}a and \ref{figure6}b to the saturation magnetization of the easy axis magnetization curves. At 300 K the value of \ha\ drops from an undoped value of 8.1 T to a value of 6.2 T for the highest achievable Ce concentration. At 400 K \ha\ drops from 5.5 T to 4.7 T for crystals with large amounts of Ce present. 

\section{Theoretical Calculations}

The $\mathrm{R_2Fe_{14}B}$ class of materials (where R = Nd, Gd, Y etc.) has been the subject of much study \cite{gu_comparative_1987, jaswal_electronic_1990, min_electronic_1993, nordstroem_calculation_1993, kitagawa_magnetic_2010}. With respect to \neo, first principles studies have been used to understand the effect of partial substitution of the Nd site by various rare earth elements (e.g., Y, La, Dy, Tb etc. \cite{liu_partitioning_2012, liu_partitioning_2013}) as well as the Fe site by Si, Ge, Sn etc.\cite{liu_fe_2014}. In this report we calculate the effects of Ce substitution for Nd on the magnetic properties of \doped, focusing on the parameters relevant for a high performance magnetic material: the saturation magnetization, Curie temperature and magnetocrystalline anisotropy. 

\subsection{Electronic and magnetic structure of \titleneo}

\begin{figure}[h]
\centering
\includegraphics[width=0.45\textwidth]{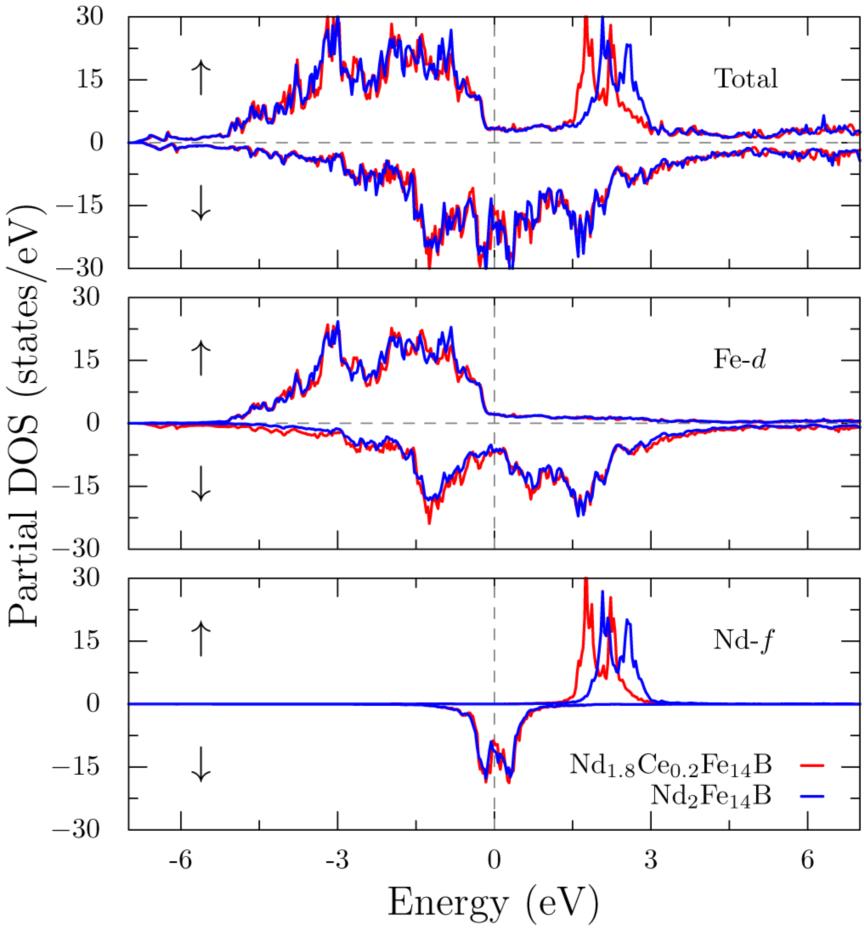}
\caption{ The total and partial density of states for undoped Nd-Fe-B (blue) and Ce doped Nd-Fe-B (red) from first-principles calculations in the ferromagnetic spin configuration. The positive and negative DOS values correspond to spin-up and spin-down channels respectively. The DOS around Fermi level is dominated by Fe-$d$ (middle panel) and Nd/Ce-$f$ (bottom panel) states.}
\label{figure7}
\end{figure}

\neo\ crystallizes in a tetragonal structure with four formula units per unit cell. In the atomic valence configuration the four electrons in the Nd-$4f$ orbitals and the six in the Fe-$3d$ orbitals make these ions strongly magnetic. The density of states (DOS) for \neo\ in the ferromagnetic state is shown in the blue curves in Fig. \ref{figure7}. Comparing the total DOS with partial DOS in the lower panels, we can see that the states near the Fermi level predominantly consist of Fe-$d$ and Nd-$f$ states. There is an exchange splitting of  ~$\sim$2 eV in the Fe-$d$ states with a nearly fully occupied spin-up channel. The spin down channel has close to two electrons, suggesting a spin magnetic moment of $\sim$3 \mub\ at the Fe sites, which is significantly above the value for bcc Fe ($\sim$2.2 \mub/Fe). For comparison, the experimental neutron diffraction data collected at 4 K shows Fe having moments between 2.370 and 3.253 \mub, within the range of these extrema.

At the Nd sites, the $4f$ states are empty in the spin-up channel and are partially occupied in the spin-down channel, showing that in this calculation the Nd moments point in the opposite direction compared to the Fe moments. The exchange splitting at the Nd-$f$ DOS is similar to that of Fe-$d$ orbitals, even though the$ f$  DOS is substantially narrower. Counting the states in the Nd-$f\downarrow$ channel up to the Fermi level gives 3.1 electrons, which suggest a spin magnetic moment of -3.1 \mub\ at the Nd sites. From our calculations, we also find an orbital moment of 2.8 \mub\ at the Nd-$f$ states (opposite to the spin moment, as expected from Hund's rules), while the orbital moments at the Fe sites are $\sim$0.06 \mub. The much larger orbital moment on the rare-earth atom suggests that the magneto-crystalline anisotropy is primarily a result of the rare-earth atom. Spin magnetic moments calculated by integrating charge density around the atom sites agree well with the above estimates. Integrating the spin-polarized charge density around the ions, we find average moments of (in \mub) -3.4,  2.5 and -0.17 for Nd, Fe and B sites respectively. The Fe moment value is larger than that for pure Fe, contributing to the excellent magnetic properties of this compound. For comparison, the experimental neutron diffraction data collected at 4 K shows Fe having net moments between 2.370 and 3.253 \mub, in good agreement with calculated value. Consistent with the experimental neutron diffraction data, we find that the Fe ions bonded with B atoms were found to have a reduced moment of 1.98 \mub. The net spin+orbital (J = S+L) magnetic moment of this system is 30.3 \mub\ per unit cell, in good agreement with both the neutron diffraction refinements and the magnetization data.

\subsection{Effects of Ce Substitution on \titleneo}

To understand the effect of Ce doping at the Nd sites on the properties of \neo, we replaced 10\% of the Nd atoms by Ce using the virtual crystal approximation (VCA). The total and partial DOS for the doped system are presented as red curves in Fig. \ref{figure7}. As we can see, the only significant change in the DOS is at the RE-$f \uparrow\  $ states, which are shifted to lower energies by $\sim$0.5 eV. Thus, we see that the electronic properties of the Nd-Fe-B magnets are not significantly affected by Ce doping, except a small reduction in the magnetic moments at the Nd/Ce site since  Ce has two fewer electrons in its valence shells ($4f^ 2$) than Nd ($4f^ 4$). Thus, the number of electrons in the 68 atom (8 rare earth atoms) unit cell is reduced by 1.6 electrons by 10\% Ce substitution which lead to a subsequent reduction in the average magnetic moment at these sites to 3.2 \mub\ from 3.4 \mub. We find from these spin-polarized calculations that the Fe moments are mostly unaffected by the substitution.

An important consequence of the reduced exchange splitting is a reduction in Curie temperature, \tc. Generally, for a ferromagnetic system characterized by large local moments such as the 2-14-1 materials, the Curie temperature is controlled by the difference in energy between the ferromagnetic ground state and an essentially antiferromagnetic structure constructed so that most of the neighbors of Fe atoms are antiparallel. This in turn is a function of the exchange parameters $J_{ij}$ connecting nearest neighbors $i$ and $j$. In the simplest mean-field approximation one may estimate the Curie temperature as a third of this energy difference, measured on a per Fe basis.

\begin{figure}[h]
\centering
\includegraphics[width=0.45\textwidth]{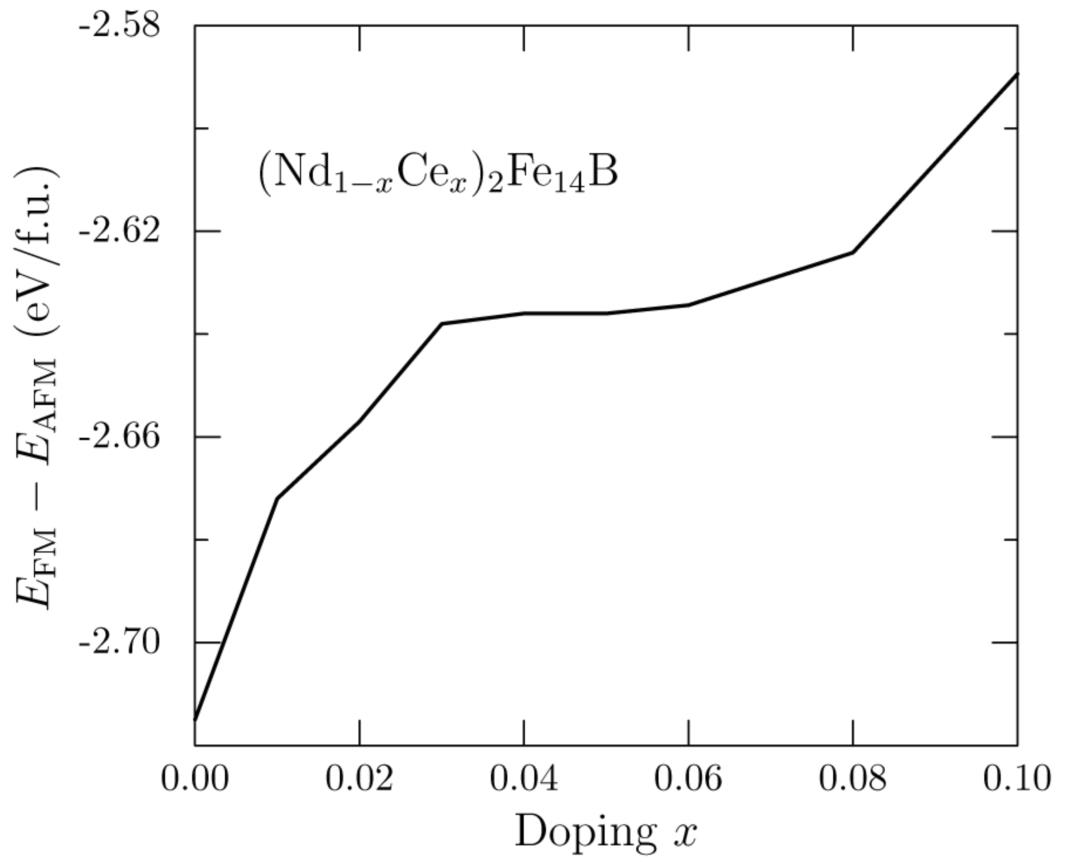}
\caption{Variation of energy difference between ferromagnetic and antiferromagnetic structures in \neo\ as function of doping parameter $x$. The energy difference reduces upon Ce doping, suggesting that the exchange interaction $J$ and consequently \tc\ will reduce as a function of Ce concentration. }
\label{figure8}
\end{figure}

We calculated the energy difference between the above-described ferromagnetic structure and an approximate antiferromagnetic structure obtained by flipping most of the nearest neighbor moments. The calculated energy difference is plotted in Fig. \ref{figure8} as a function of the Ce concentration$x$. The $x = 0$ value of -2.71 eV per formula unit corresponds in the mean-field approximation to a \tc\ of 748 K, which is to be compared to the experimental value of 585 K. The smaller experimental value suggests that effects beyond our simple mean-field approach may be relevant. A similar calculation overestimate was also found in Ref. \cite{alam_site-preference_2013}. Upon substituting Ce atoms via VCA, the magnitude of the energy difference decreases by about 4 percent for a 10 percent Ce substitution. This suggests that the mean field $J$ decreases in \neo\ as $x$ is increased which would in turn decrease the Curie temperature. This is consistent with our experimental results, which also find a decrease in Curie temperature with Ce alloying.

Finally, we calculated the magnetic anisotropy energy (MAE) of \doped\ for Ce concentrations $ x = 0$ and  $x = 0.1$  to understand the effect of doping on this important parameter. We used the experimental structure at room temperature (i.e., in the ferromagnetic state) to perform these calculations \cite{shoemaker_structure_1984}. The total energy is then calculated in the presence of spin orbit coupling with the moments pointing along $a$ and $c$ directions. The MAE is defined as $K_1 = E_a - E_c$. We find that for $x = 0$, $K_1$ = 12.3 meV/f.u. ($\sim$8 MJ/m$^3$). This is somewhat larger than the experimental value of 4.3 MJ/m$^3$, but we note that temperature effects are known to be significant in this compound \cite{goll_high-performance_2000}.  We find that the anisotropy energy remains more or less same upon doping; we get $K_1$ = 13.2 meV/f.u. when $x = 0.1$. We do not consider the increase to be significant, but expect that larger levels of Ce alloying would tend to show a reduction in the value of the MAE. In short, we find that that the high magnetic anisotropy energy of the Nd-Fe-B system is not significantly affected by the Ce doping, which confirms experimental observations.

\section{Summary}

Large single crystals of \neo\ were grown from Nd-Fe flux. These crystals were a maximum of $\sim$0.5 g in mass and displayed sharp faceting. Cerium substitution was attempted for Nd; the maximum degree of substitution we were able to achieve was the composition \bigneo. We found that Ce and Nd formed a continuous solid solution in contrast to the melt-spun ribbons of similar compositions fabricated by Pathak $et$ $al$. \cite{pathak_cerium:_2015} where a two phase region was observed for $ 0.15 < x < 0.4 $ in \doped. Single crystal neutron diffraction showed that the Ce has a slight preference for the larger RE $4g$ site, favoring this location by a factor of $\sim$7:3 over the RE $4f$ site.

The presence of Ce serves to deleteriously affect magnetic properties such as \tc, \ms, and \ha. The values of these characteristics largely fall in line with empirical predictions based on the application of simple alloying rules between the two parent compounds (\neo\ and $\mathrm{Ce_2Fe_{14}B}$). 

We have also used X-ray photoelectron spectroscopy and synchrotron emissions to determine the oxidation states and binding energies of all species in Ce-substituted \neo; preliminary data shows that the XPS spectrum for the Ce 3$d$ electrons most closely resembles that of $\mathrm{Ce_2O_3}$.

With the help of first-principles calculations, we studied the electronic and magnetic properties of Ce doped Nd-Fe-B magnets. We find that for up to $ x = 0.1$ doping of Ce, the electronic structure around the Fermi level is not significantly affected. The exchange splitting and the magnetic moments at the Nd sites are diminished as a consequence of Ce doping, which in turn reduces the Curie temperature. Finally, we find that the high magnetic anisotropy energy of the Nd-Fe-B system is not significantly altered by the Ce doping, confirming the experimental observations.

\begin{table*}
  \centering
  \caption{Summary of 300 K single crystal neutron diffraction data of sample \neuneo}
    \begin{tabular}{llccccc}
& & & & & & \\   
    \multicolumn{1}{c}{} & \multicolumn{1}{c}{} & \multicolumn{3}{c}{\textbf{Atomic Coordinates}} & \multicolumn{1}{c}{}  & \multicolumn{1}{c}{} \\ 
	    \multicolumn{1}{c}{\textbf{Lattice Constants}} & \multicolumn{1}{c}{\textbf{Site}} & \multicolumn{1}{c}{\textit{x}} & \multicolumn{1}{c}{\textit{y}} & \multicolumn{1}{c}{\textit{z}} & \multicolumn{1}{c}{$\bm {U_{iso}} (\AA^2)$} & \multicolumn{1}{c}{\textbf{Occ. f.}} \\
\cline{3-5} \\
    $a = b = 8.8032 \AA$  & B1  ($4g$) & \multicolumn{1}{c}{0.12430(33)} & \multicolumn{1}{c}{0.12430(33)} & \multicolumn{1}{c}{0} & \multicolumn{1}{c}{0.0110(8)} & \multicolumn{1}{c}{1} \\
    $c = 12.1880 \AA$ & Fe1 ($16k_1$) & \multicolumn{1}{c}{0.03749(14)} & \multicolumn{1}{c}{0.35957(13)} & \multicolumn{1}{c}{0.32421(9)} & \multicolumn{1}{c}{0.0081(3)} & \multicolumn{1}{c}{1} \\
    $\alpha=\beta=\gamma=90^{\circ}$ & Fe2 ($16k_2$) & \multicolumn{1}{c}{0.06708(14)} & \multicolumn{1}{c}{0.27577(14)} & \multicolumn{1}{c}{0.12752(9)} & \multicolumn{1}{c}{0.0081(3)} & \multicolumn{1}{c}{1} \\
          & Fe3 ($8j_1$) & \multicolumn{1}{c}{0.09785(14)} & \multicolumn{1}{c}{0.09785(14)} & \multicolumn{1}{c}{0.29589(13)} & \multicolumn{1}{c}{0.0081(3)} & \multicolumn{1}{c}{1} \\
    $P4_2/mnm$ & Fe4 ($8j_2$) & \multicolumn{1}{c}{0.31749(14)} & \multicolumn{1}{c}{0.31749(14)} & \multicolumn{1}{c}{0.25435(13)} & \multicolumn{1}{c}{0.0081(3)} & \multicolumn{1}{c}{1} \\
          & Fe5 ($4e$) & \multicolumn{1}{c}{0} & \multicolumn{1}{c}{0} & \multicolumn{1}{c}{0.11502(20)} & \multicolumn{1}{c}{0.0081(3)} & \multicolumn{1}{c}{1} \\
          & Fe6 ($4c$) & \multicolumn{1}{c}{0} & \multicolumn{1}{c}{½} & \multicolumn{1}{c}{0} & \multicolumn{1}{c}{0.0081(3)} & \multicolumn{1}{c}{1} \\
          & Nd1 ($4g$) & \multicolumn{1}{c}{0.23023(27)} & \multicolumn{1}{c}{0.76977(27)} & \multicolumn{1}{c}{0} & \multicolumn{1}{c}{0.0079(5)} & \multicolumn{1}{c}{0.69} \\
          & Ce1 ($4g$) & \multicolumn{1}{c}{0.23023(27)} & \multicolumn{1}{c}{0.76977(27)} & \multicolumn{1}{c}{0} & \multicolumn{1}{c}{0.0079(5)} & \multicolumn{1}{c}{0.31} \\
          & Nd2 ($4f$) & \multicolumn{1}{c}{0.35694(27)} & \multicolumn{1}{c}{0.35694(27)} & \multicolumn{1}{c}{0} & \multicolumn{1}{c}{0.0079(5)} & \multicolumn{1}{c}{0.87} \\
          & Ce2 ($4f$) & \multicolumn{1}{c}{0.35694(27)} & \multicolumn{1}{c}{0.35694(27)} & \multicolumn{1}{c}{0} & \multicolumn{1}{c}{0.0079(5)} & \multicolumn{1}{c}{0.13} \\
          &       & \multicolumn{1}{c}{} & \multicolumn{1}{c}{} & \multicolumn{1}{c}{} & \multicolumn{1}{c}{} & \multicolumn{1}{c}{} \\
          &       & \multicolumn{3}{c}{\textbf{Magnetic vectors}} & \multicolumn{1}{c}{\multirow{2}[3]{*}{}} &  \\
          &       & r ($\mu_{B}$) & \textit{$\phi$} & \textit{$\theta$} & \multicolumn{1}{c}{} &  \\
\cline{3-5} \\
    \textbf{Total Moment} & Fe1 ($16k_1$) & 2.869(137) & 0     & 0     & \multicolumn{2}{r}{} \\
    $36.76 \pm 2.32 \  \mu_{B}$ & Fe2 ($16k_2$) & 2.247(135) & 0     & 0     & \multicolumn{2}{r}{} \\
          & Fe3 ($8j_1$) & 1.614(169) & 0     & 0     & \multicolumn{2}{r}{} \\
          & Fe4 ($8j_2$) & 2.268(182) & 0     & 0     & \multicolumn{2}{r}{} \\
          & Fe5 ($4e$) & 3.247(159) & 0     & 0     & \multicolumn{2}{r}{} \\
          & Fe6 ($4c$) & 1.305(137) & 0     & 0     & \multicolumn{2}{r}{} \\
          & Nd1 ($4g$) & 2.372(119) & 0     & 0     & \multicolumn{2}{r}{} \\
          & Ce1 ($4g$) & 2.372(119) & 0     & 0     & \multicolumn{2}{r}{} \\
          & Nd2 ($4f$) & 1.605(118) & 0     & 0     & \multicolumn{2}{r}{} \\
          & Ce2 ($4f$) & 1.605(118) & 0     & 0     & \multicolumn{2}{r}{} \\
          &       &       &       &       &       &  \\
         \cline{1-7} \\
    \textbf{Reliability factors} &       &       &       &       &       &  \\
          & $R_{F2}$ & $w_{F2}$ & $R_F$ & $\chi^2$ &       &  \\
          & 0.0473 & 0.0775 & 0.0461 & 4.06  &       &  \\
          &       &       &       &       &       &  \\  
    \end{tabular}%
  \label{table4}%
\end{table*}%

\begin{table*}
  \centering
  \caption{Summary of 4 K single crystal neutron diffraction data of sample \neuneo}
    \begin{tabular}{llccccc}
& & & & & & \\   
    \multicolumn{1}{c}{} & \multicolumn{1}{c}{} & \multicolumn{3}{c}{\textbf{Atomic Coordinates}} & \multicolumn{1}{c}{}  & \multicolumn{1}{c}{} \\ 
	    \multicolumn{1}{c}{\textbf{Lattice Constants}} & \multicolumn{1}{c}{\textbf{Site}} & \multicolumn{1}{c}{\textit{x}} & \multicolumn{1}{c}{\textit{y}} & \multicolumn{1}{c}{\textit{z}} & \multicolumn{1}{c}{$\bm {U_{iso}} (\AA^2)$} & \multicolumn{1}{c}{\textbf{Occ. f.}} \\
\cline{3-5} \\
   $a = b = 8.8000 \AA$& B1  ($4g$) & \multicolumn{1}{c}{0.12377(41)} & \multicolumn{1}{c}{0.12377(41)} & \multicolumn{1}{c}{0} & \multicolumn{1}{c}{0.0051(11)} & \multicolumn{1}{c}{1} \\
    $c = 12.1880 \AA$ & Fe1 ($16k_1$) & \multicolumn{1}{c}{0.03736(16)} & \multicolumn{1}{c}{0.36009(16)} & \multicolumn{1}{c}{0.32426(11)} & \multicolumn{1}{c}{0.0018(4)} & \multicolumn{1}{c}{1} \\
    $\alpha=\beta=\gamma=90^{\circ}$ & Fe2 ($16k_2$) & \multicolumn{1}{c}{0.06698(17)} & \multicolumn{1}{c}{0.27540(17)} & \multicolumn{1}{c}{0.12759(11)} & \multicolumn{1}{c}{0.0018(4)} & \multicolumn{1}{c}{1} \\
          & Fe3 ($8j_1$) & \multicolumn{1}{c}{0.09795(17)} & \multicolumn{1}{c}{0.09795(17)} & \multicolumn{1}{c}{0.29655(17)} & \multicolumn{1}{c}{0.0018(4)} & \multicolumn{1}{c}{1} \\
     $P4_2/mnm$& Fe4 ($8j_2$) & \multicolumn{1}{c}{0.31775(17)} & \multicolumn{1}{c}{0.31775(17)} & \multicolumn{1}{c}{0.25415(16)} & \multicolumn{1}{c}{0.0018(4)} & \multicolumn{1}{c}{1} \\
          & Fe5 ($4e$) & \multicolumn{1}{c}{0} & \multicolumn{1}{c}{0} & \multicolumn{1}{c}{0.11597(24)} & \multicolumn{1}{c}{0.0018(4)} & \multicolumn{1}{c}{1} \\
          & Fe6 ($4c$) & \multicolumn{1}{c}{0} & \multicolumn{1}{c}{½} & \multicolumn{1}{c}{0} & \multicolumn{1}{c}{0.0018(4)} & \multicolumn{1}{c}{1} \\
          & Nd1 ($4g$) & \multicolumn{1}{c}{0.22988(28)} & \multicolumn{1}{c}{0.77012(28)} & \multicolumn{1}{c}{0} & \multicolumn{1}{c}{0.0009(5)} & \multicolumn{1}{c}{0.69} \\
          & Ce1 ($4g$) & \multicolumn{1}{c}{0.22988(28)} & \multicolumn{1}{c}{0.77012(28)} & \multicolumn{1}{c}{0} & \multicolumn{1}{c}{0.0009(5)} & \multicolumn{1}{c}{0.31} \\
          & Nd2 ($4f$) & \multicolumn{1}{c}{0.35713(27)} & \multicolumn{1}{c}{0.35713(27)} & \multicolumn{1}{c}{0} & \multicolumn{1}{c}{0.0009(5)} & \multicolumn{1}{c}{0.87} \\
          & Ce2 ($4f$) & \multicolumn{1}{c}{0.35713(27)} & \multicolumn{1}{c}{0.35713(27)} & \multicolumn{1}{c}{0} & \multicolumn{1}{c}{0.0009(5)} & \multicolumn{1}{c}{0.13} \\
          &       & \multicolumn{1}{c}{} & \multicolumn{1}{c}{} & \multicolumn{1}{c}{} & \multicolumn{1}{c}{} & \multicolumn{1}{c}{} \\
          &       & \multicolumn{3}{c}{\textbf{Magnetic vectors}} & \multicolumn{1}{c}{\multirow{2}[3]{*}{}} &  \\
             &       & r ($\mu_{B}$) & \textit{$\phi$} & \textit{$\theta$} & \multicolumn{1}{c}{} &  \\
\cline{3-5} \\
   \textbf{ Total Moment} & Fe1 ($16k_1$) & 3.288(76) & 226.09(569) & 39.00(180) & \multicolumn{2}{r}{} \\
   $36.76 \pm 2.32 \ \mu_{B}$ & Fe2 ($16k_2$) & 2.878(76) & 199.04(510) & 39.00(180) & \multicolumn{2}{r}{} \\
& Fe3 ($8j_1$) & 3.378(115) & 156.65(527) & 39.00(180) & \multicolumn{2}{r}{} \\
          & Fe4 ($8j_2$) & 3.528(92) & 191.39(471) & 39.00(180) & \multicolumn{2}{r}{} \\
          & Fe5 ($4e$) & 3.302(103) & 240.22(706) & 39.00(180) & \multicolumn{2}{r}{} \\
          & Fe6 ($4c$) & 2.819(131) & 210.84(711) & 39.00(180) & \multicolumn{2}{r}{} \\
          & Nd1 ($4g$) & 2.921(70) & 255.63(732) & 39.00(180) & \multicolumn{2}{r}{} \\
          & Ce1 ($4g$) & 2.921(70) & 255.63(732) & 39.00(180) & \multicolumn{2}{r}{} \\
          & Nd2 ($4f$) & 3.644(96) & 185.29(536) & 39.00(180) & \multicolumn{2}{r}{} \\
          & Ce2 ($4f$) & 3.644(96) & 185.29(536) & 39.00(180) & \multicolumn{2}{r}{} \\
          &       &       &       &       &       &  \\
       \cline{1-7} \\
    \textbf{Reliability factors} &       &       &       &       &       &  \\
        & $R_{F2}$ & $w_{F2}$ & $R_F$ & $\chi^2$ &       &   \\
          & 0.0459 & 0.0948 & 0.0648 & 0.0738 &       &  \\   
    \end{tabular}%
  \label{table5}%
\end{table*}%


\begin{thebibliography}{10}

\bibitem{_critical_2011}
{Critical Materials Strategy}.
\newblock Technical report, United States Department of Energy, December 2011.

\bibitem{sagawa_permanent_1984}
M.~Sagawa, S.~Fujimura, H.~Yamamoto, Y.~Matsuura, and K.~Hiraga.
\newblock {Permanent Magnet Materials Based on the Rare Earth-Iron-Boron
  Tetragonal Compounds.}
\newblock {\em IEEE Transactions on Magnetics}, MAG-20:1584--9, 1984.

\bibitem{croat_praseodymium-iron-_1984}
J.~J. Croat, J.~F. Herbst, R.~W. Lee, and F.~E. Pinkerton.
\newblock {Praseodymium-Iron- and Neodymium-Iron-Based Materials: A New Class
  of High-Performance Permanent Magnets.}
\newblock {\em Journal of Applied Physics}, 55:2078--82, 1984.

\bibitem{pathak_cerium:_2015}
A.~K. Pathak, M.~Khan, .~A. Gschneidner, Jr., R.~W. McCallum, L.~Zhou, K.~Sun,
  K.~W. Dennis, C.~Zhou, F.~E. Pinkerton, M.~J. Kramer, and V.~K. Pecharsky.
\newblock Cerium: {An} {Unlikely} {Replacement} of {Dysprosium} in {High}
  {Performance} {Nd}-{Fe}-{B} {Permanent} {Magnets}.
\newblock {\em Advanced Materials}, page Ahead of Print, 2015.

\bibitem{givord_magnetic_1984}
D.~Givord, H.~S. Li, and J.~M. Moreau.
\newblock {Magnetic Properties and Crystal Structure of Neodymium-Iron-Boron
  ({Nd}$_{\textrm{2}}${Fe}$_{\textrm{14}}${B}).}
\newblock {\em Solid State Communications}, 50:497--9, 1984.

\bibitem{herbst_relationships_1984}
J.~F. Herbst, J.~J. Croat, F.~E. Pinkerton, and W.~B. Yelon.
\newblock {Relationships Between Crystal Structure and Magnetic Properties in
  Neodymium-Iron-Boron ({Nd}$_{\textrm{2}}${Fe}$_{\textrm{14}}${B}).}
\newblock {\em Physical Review B}, 29:4176--8, 1984.

\bibitem{herbst_rare_1991}
J.~F. Herbst.
\newblock {Rare Earth({R}) Iron Boron
  ({R}$_{\textrm{2}}${Fe}$_{\textrm{14}}${B}) Materials: Intrinsic Properties
  and Technological Aspects.}
\newblock {\em Reviews of Modern Physics}, 63:819--98, 1991.

\bibitem{abache_magnetic_1986}
C.~Abache and J.~Oesterreicher.
\newblock {Magnetic Anisotropies and Spin Reorientations of
  {R}$_{\textrm{2}}${Fe}$_{\textrm{14}}${B}-Type Compounds.}
\newblock {\em Journal of Applied Physics}, 60:3671--9, 1986.

\bibitem{alam_site-preference_2013}
A.~Alam, M.~Khan, R.~W. McCallum, and D.~D. Johnson.
\newblock {Site-Preference and Valency for Rare-Earth Sites in
  ({R}-{Ce})$_{\textrm{2}}${Fe}$_{\textrm{14}}${B} Magnets.}
\newblock {\em Applied Physics Letters}, 102:042402/1--042402/4, 2013.

\bibitem{canfield_high-temperature_2001}
P.~C. Canfield and I.~R. Fisher.
\newblock {High-Temperature Solution Growth of Intermetallic Single Crystals
  and Quasicrystals.}
\newblock {\em Journal of Crystal Growth}, 225:155--161, 2001.

\bibitem{canfield_private_2014}
P.~C. Canfield.
\newblock {Private Communication}, 2014.

\bibitem{lewis_magnetic_1998}
L.~H. Lewis, J.-Y. Wang, and P.~Canfield.
\newblock {Magnetic Domains of Single-Crystal
  {Nd}$_{\textrm{2}}${Fe}$_{\textrm{14}}${B} Imaged by Unmodified Scanning
  Electron Microscopy.}
\newblock {\em Journal of Applied Physics}, 83:6843--6845, 1998.

\bibitem{saparov_private_2014}
B.~Saparov.
\newblock Private {Communication}, 2014.

\bibitem{sheldrick:sc5010}
George~M. Sheldrick.
\newblock {{A Short History of {\it SHELX}}}.
\newblock {\em Acta Crystallographica Section A}, 64(1):112--122, Jan 2008.

\bibitem{Chako_2011}
B.~C. Chakoumakos, H.~Cao, F.~Ye, A.~D. Stoica, M.~Popovici, M.~Sundaram,
  W.~Zhou, J.~S. Hicks, G.~W. Lynn, and R.~A. Riedel.
\newblock {Four-Circle Single-Crystal Neutron Diffractometer at the High Flux
  Isotope Reactor}.
\newblock {\em Journal of Applied Crystallography}, 44(3):655--658, 2011.

\bibitem{carvajal_1993}
J.~Rodríguez-Carvajal.
\newblock {Recent Advances in Magnetic Structure Determination by Neutron
  Powder Diffraction.}
\newblock {\em Physica B}, 192(1–2):55 -- 69, 1993.

\bibitem{axnanda_using_2015}
S.~Axnanda, E.~J. Crumlin, S.~Rani, Z.~Hussain, B.~Mao, R.~Chang, P.~G.
  Karlsson, Marten O.~M. E., M.~Lundqvist, R.~Moberg, P.~Ross, and Z.~Liu.
\newblock Using "{Tender}" {X}-ray {Ambient} {Pressure} {X}-{Ray}
  {Photoelectron} {Spectroscopy} as {A} {Direct} {Probe} of {Solid}-{Liquid}
  {Interface}.
\newblock {\em Scientific Reports}, 5:9788, 2015.

\bibitem{Perdew_1996}
J.~P. Perdew, K.~Burke, and M.~Ernzerhof.
\newblock {Generalized Gradient Approximation Made Simple}.
\newblock {\em Phys. Rev. Lett.}, 77:3865--3868, Oct 1996.

\bibitem{bloechl_projector_1994}
P.~E. Bloechl.
\newblock {Projector Augmented-Wave Method.}
\newblock {\em Physical Review B}, 50:17953--79, 1994.

\bibitem{Kresse_ab_1993}
G.~Kresse and J.~Hafner.
\newblock {Ab Initio Molecular Dynamics of Liquid Metals.}
\newblock {\em Physical Review B}, 47:558--61, 1993.

\bibitem{Kresse_efficient_1996}
G.~Kresse and J.~Furthmueller.
\newblock {Efficient Iterative Schemes for Ab Initio Total-Energy Calculations
  Using a Plane-Wave Basis Set.}
\newblock {\em Physical Review B}, 54:11169--11186, 1996.

\bibitem{abache_magnetic_1985}
C.~Abache and H.~Oesterreicher.
\newblock {Magnetic Properties of Compounds Rare Earth-Iron-Boron
  ({R}$_{\textrm{2}}${Fe}$_{\textrm{14}}${B}).}
\newblock {\em Journal of Applied Physics}, 57:4112--14, 1985.

\bibitem{herbst_preferential_1986}
J.~F. Herbst and W.~B. Yelon.
\newblock {Preferential Site Occupation and Magnetic Structure of
  {Nd}$_{\textrm{2}}$({Co}$_{\textrm{x}}${Fe}$_{\textrm{(1-x)}}$)$_{\textrm{14}}${B}
  Systems}.
\newblock {\em Journal of Applied Physics}, 60(12), 1986.

\bibitem{dong_2015}
G.~Dong, Y.~Sui, P.~Qian, Y.~Wu, and L.~Guo.
\newblock {Experimental and Theoretical Studies on Site Preference of Ti in
  $\mathrm{Nd_2(Fe,Ti)_{14}B}$}.
\newblock {\em Journal of Magnetism and Magnetic Materials}, 379:108 -- 111,
  2015.

\bibitem{teplykh_2013}
A.E. Teplykh, Yu.G. Chukalkin, S.~Lee, S.G. Bogdanov, N.V. Kudrevatykh, E.V.
  Rosenfeld, Yu.N. Skryabin, Y.~Choi, A.V. Andreev, and A.N. Pirogov.
\newblock {Magnetism of Ordered and Disordered Alloys of $\mathrm{R_2Fe_{14}B}$
  (R = Nd, Er) Type}.
\newblock {\em Journal of Alloys and Compounds}, 581:423 -- 430, 2013.

\bibitem{givord_polarized_1985}
D.~Givord, H.~S. Li, and F.~Tasset.
\newblock {Polarized Neutron Study of the Compounds Yttrium-Iron-Boron
  ({Y}$_{\textrm{2}}${Fe}$_{\textrm{14}}${B}) and Neodymium-Iron-Boron
  ({Nd}$_{\textrm{2}}${Fe}$_{\textrm{14}}${B}).}
\newblock {\em Journal of Applied Physics}, 57:4100--2, 1985.

\bibitem{hirosawa_magnetization_1986}
S.~Hirosawa, Y.~Matsuura, H.~Yamamoto, S.~Fujimura, M.~Sagawa, and H.~Yamauchi.
\newblock {Magnetization and Magnetic Anisotropy of
  {R}$_{\textrm{2}}${Fe}$_{\textrm{14}}${B} Measured on Single Crystals.}
\newblock {\em Journal of Applied Physics}, 59:873--9, 1986.

\bibitem{gu_comparative_1987}
Z.~Gu and W.~Y. Ching.
\newblock {Comparative Studies of Electronic and Magnetic Structures in Yttrium
  Iron Boride ({Y}$_{\textrm{2}}${Fe}$_{\textrm{14}}${B}), Neodymium Iron
  Boride ({Nd}$_{\textrm{2}}${Fe}$_{\textrm{14}}${B}), Yttrium Cobalt Boride
  ({Y}$_{\textrm{2}}${Co}$_{\textrm{14}}${B}), and Neodymium Cobalt Boride
  ({Nd}$_{\textrm{2}}${Co}$_{\textrm{14}}${B}).}
\newblock {\em Physical Review B}, 36:8530--46, 1987.

\bibitem{jaswal_electronic_1990}
S.~S. Jaswal.
\newblock {Electronic Structure and Magnetism of
  {R}$_{\textrm{2}}${Fe}$_{\textrm{14}}${B} ({R} = Yttrium, Neodymium)
  Compounds.}
\newblock {\em Physical Review B}, 41:9697--700, 1990.

\bibitem{min_electronic_1993}
B.~I. Min, J.~S. Kang, J.~H. Hong, J.~I. Jeong, Y.~P. Lee, S.~D. Choi, W.~Y.
  Lee, C.~J. Yang, and C.~G. Olson.
\newblock {Electronic and Magnetic Structures of the Rare-Earth Permanent
  Magnet Neodymium Iron Boride ({Nd}$_{\textrm{2}}${Fe}$_{\textrm{14}}${B}).}
\newblock {\em Physical Review B}, 48:6217--24, 1993.

\bibitem{nordstroem_calculation_1993}
L.~Nordstroem, B.~Johansson, and M.~S.~S. Brooks.
\newblock {Calculation of the Electronic Structure and the Magnetic Moments of
  Neodymium Iron Boride ({Nd}$_{\textrm{2}}${Fe}$_{\textrm{14}}${B}).}
\newblock {\em Journal of Physics: Condensed Matter}, 5:7859--70, 1993.

\bibitem{kitagawa_magnetic_2010}
I.~Kitagawa and Y.~Asari.
\newblock {Magnetic Anisotropy of {R}$_{\textrm{2}}${Fe}$_{\textrm{14}}${B}
  ({R}={Nd}, {Gd}, {Y}): {Density} Functional Calculation by Using the Linear
  Combination of Pseudo-Atomic-Orbital Method.}
\newblock {\em Physical Review B}, 81:214408/1--214408/7, 2010.

\bibitem{liu_partitioning_2012}
X.~B. Liu and Z.~Altounian.
\newblock {The Partitioning of {Dy} and {Tb} in {NdFeB} Magnets: {A}
  First-Principles Study.}
\newblock {\em Journal of Applied Physics}, 111:07A701/1--07A701/3, 2012.

\bibitem{liu_partitioning_2013}
X.~B. Liu, Z.~Altounian, M.~Huang, Q.~Zhang, and J.~P.. Liu.
\newblock {The Partitioning of {La} and {Y} in {Nd}-{Fe}-{B} Magnets: {A}
  First-Principles Study.}
\newblock {\em Journal of Alloys and Compounds}, 549:366--369, 2013.

\bibitem{liu_fe_2014}
X.~B. Liu, J.~P. Liu, Q.~Zhang, and Z.~Altounian.
\newblock {The {Fe} Substitution in
  {Nd}$_{\textrm{2}}$({Fe},{M})$_{\textrm{14}}${B} ({M} = {Si}, {Ge} and {Sn}):
  {A} First-Principles Study.}
\newblock {\em Computational Materials Science}, 85:186--192, 2014.

\bibitem{shoemaker_structure_1984}
C.~B. Shoemaker, D.~P. Shoemaker, and R.~Fruchart.
\newblock {The Structure of a New Magnetic Phase Related to the Sigma Phase:
  Iron Neodymium Boride {Nd}$_{\textrm{2}}${Fe}$_{\textrm{14}}${B}.}
\newblock {\em Acta Crystallographica, Section C}, C40:1665--8, 1984.

\bibitem{goll_high-performance_2000}
D.~Goll and H.~Kronmuller.
\newblock {High-Performance Permanent Magnets.}
\newblock {\em Naturwissenschaften}, 87:423--438, 2000.

\end{thebibliography}
\end{document}